\documentclass[preprintnumbers,superscriptaddress,nofootinbib,aps,preprint, floatfix,longbibliography]{revtex4-1}
\usepackage[top=1in, bottom=1in, left=1.in, right=1.in]{geometry}
\usepackage[english]{babel}
\usepackage{atbegshi}
\usepackage{amsmath,amssymb}
\usepackage{amsbsy,amstext}
\usepackage{amsthm, simplewick,mathrsfs}
\usepackage{graphicx}
\usepackage{amsfonts}
\usepackage{upgreek}
\usepackage[titletoc]{appendix}
\usepackage[dvipsnames]{xcolor}
\usepackage{slashed}
\usepackage{comment}
\usepackage[colorlinks=true,urlcolor=blue
,anchorcolor=blue,citecolor=blue,filecolor=blue,linkcolor=blue,menucolor=blue
]{hyperref}

\newcommand{\MeV}{\mathrm{MeV}}
\newcommand{\eV}{\mathrm{eV}}

\newcommand{\TRH}{T_\mathrm{RH}}

\newcommand{\beq}{\begin{equation}}
\newcommand{\eeq}{\end{equation}}

\newcommand{\arh}{a_{\mathrm{RH}}}
\newcommand{\ahor}{a_{\mathrm{hor}}}
\newcommand{\aeq}{a_{\mathrm{eq}}}
\newcommand{\keq}{k_{\mathrm{eq}}}
\newcommand{\zeq}{z_{\mathrm{eq}}}

\begin{document}
\title{Dark Matter Microhalos From Simplified Models}

\author{Nikita Blinov}
\affiliation{Fermi National Accelerator Laboratory, Batavia, IL 60510, USA\\[0.1cm]}
\affiliation{Kavli Institute for Cosmological Physics, University of Chicago, Chicago, IL 60637, USA\\[0.1cm]}

\author{Matthew J. Dolan}
\affiliation{ARC Centre of Excellence for Dark Matter Particle Physics, School of Physics, University of Melbourne, 3010, Australia\\[0.1cm]}

\author{Patrick Draper}
\affiliation{Illinois Center for Advanced Studies of the Universe, Department of Physics, University of Illinois, Urbana, IL 61801, USA\\[0.1cm]}

\author{Jessie Shelton}
\affiliation{Illinois Center for Advanced Studies of the Universe, Department of Physics, University of Illinois, Urbana, IL 61801, USA\\[0.1cm]}

\begin{abstract}
We introduce simplified models for enhancements in the matter power spectrum at small scales and study their implications for dark matter substructure and gravitational observables. These models capture the salient aspects of a variety of early universe scenarios that predict enhanced small-scale structure, such as axion-like particle dark matter, light vector dark matter, and epochs of early matter domination. We use a model-independent, semi-analytic treatment to map bumps in the matter power spectrum to early-forming sub-solar mass dark matter halos and estimate their  evolution, disruption, and contribution to substructure of clusters and galaxies at late times. We discuss the sensitivity of gravitational observables, including pulsar timing arrays and caustic microlensing, to both the presence of bumps in the power spectrum  and variations in their basic properties.

\end{abstract}

\preprint{FERMILAB-PUB-21-015-AE-T}
\date{\today}

\maketitle
\tableofcontents

\section{Introduction}
\label{sec:intro}

Many different microphysical models can produce features in the matter power spectrum (MPS) in the early universe. These features correspond to enhancement or suppression of the MPS on given scales relative to the predictions of the minimal Cold Dark Matter (CDM) scenario. Sufficiently pronounced deviations on large enough scales can lead to observable signatures in the late-time matter distribution.

In this paper, we study simplified models for modifications of the MPS on scales well below the baryon Jeans length. The resulting dark matter (DM) substructure (which we refer to interchangeably as microhalos, minihalos or simply ``clumps") form at high redshift; as a result, they are significantly less massive than small-scale structure often considered in the literature (e.g., dwarf and satellite galaxies up to $10^{10}M_{\odot}$ in mass). Instead, our focus is on \emph{very} small-scale, non-compact structures orders of magnitude lighter than our Sun, which do not host baryonic objects such as stars.

We consider a range of localized features in the small-scale MPS 
and study the resulting observational signatures in the late-time universe; we focus on signatures that depend only on the gravitational properties of the DM distribution.  Our aim is to develop a simple procedure to approximately predict the properties of small-scale DM substructure from the broad features of an underlying MPS, and to illustrate the sensitivity of gravitational observation techniques in a model-independent way.

Microphysical models that yield bump-like features can be classified according to whether fluctuations in the density contrast begin with small initial amplitudes, and features are created or magnified over time in the process of cosmic expansion, or whether fluctuations begin with large initial amplitudes. 
Example models in the first class include standard dark matter models with a modified expansion history such that primordial matter perturbations experience a period of enhanced growth. Studies of models of this type have addressed the impact on small scale structure for WIMP DM candidates~\cite{Erickcek:2011us,Barenboim:2013gya,Fan:2014zua}, ALPs/axions~\cite{Visinelli:2018wza,Nelson:2018via,Blinov:2019jqc}, hidden sector DM \cite{Zhang:2015era,Dror:2016rxc,Dror:2017gjq,Blanco:2019eij,Erickcek:2020wzd}, and others.
Models in the second class do not rely on large alterations to the expansion history. Examples include vector DM from inflationary fluctuations~\cite{Graham:2015rva,Ema:2019yrd,Ahmed:2020fhc,Kolb:2020fwh}, ALPs/axions with post-inflationary $U(1)_{PQ}$ breaking~\cite{Kolb:1994fi,Fairbairn:2017sil,Vaquero:2018tib,Buschmann:2019icd}, ALPs/axions with parametric resonance enhancement~\cite{Arvanitaki:2019rax}, large semi-diffuse solitons, and others.

Given this profusion of models that lead to similar phenomenology in  small-scale structure, we argue for a simplified model-style approach. This approach allows us to abstract away the microphysics and detailed dependence on any specific dark matter model, and study instead a series of simple, few-parameter modifications to the matter power spectrum which efficiently and accurately capture the dominant physical effects of a broad range of models.

One can contemplate many different modifications to the MPS. From the microphysical point of view, one of the simplest features to generate is a {\em bump}. A bump is characterized by a rise in the power spectrum over a range of wavenumbers $k$, followed by a peak (which may be narrow or broad), and finally a short-distance cutoff at higher $k$. The properties of both the rising (smaller $k$) and falling (larger $k$) sides of the bump  are  ultimately determined by microphysics. For example, in models where a bump is generated by a period of modified cosmic expansion prior to Big Bang Nucleosynthesis (BBN), the rising slope is determined by the equation of state of the universe during the epoch of modified expansion. Both a period of early matter domination (EMD) and kination lead to linear growth in density perturbations~\cite{Erickcek:2011us,Barenboim:2013gya,Fan:2014zua,Zhang:2015era,Erickcek:2015jza,Redmond:2018xty,Blanco:2019eij,Blinov:2019jqc,Erickcek:2020wzd,Kadota:2020ahr}. However, EMD leads to a rise in the (dimensionless) power spectrum that goes like $k^4$, while for kination the power spectrum grows like $k$. The largest scale that inherits this enhanced growth---i.e., the location of the bump---is determined by other aspects of the microphysics, for example, the scale of reheating. On the small-scale side, the short-distance cutoff can be associated with a free-streaming scale, a Jeans length, a kinetic decoupling scale, cannibal oscillations, and so on.

Observational signatures can be divided according to whether they are sensitive primarily to the gravitational properties of DM or its other, model-dependent interactions. Examples of non-gravitational search techniques sensitive to very small-scale DM substructure include direct and indirect detection experiments. In particular, a number of works have explored the possibility that enhanced DM substructure might boost the sensitivity of indirect searches for DM annihilations~\cite{Blanco:2019eij,Kadota:2020ahr}. Examples of purely gravitational signatures include stellar~\cite{Croon:2020wpr,Croon:2020ouk} and caustic~\cite{Dai:2019lud} microlensing, and pulsar timing measurements~\cite{Siegel:2007fz,Baghram:2011is,Kashiyama:2018gsh,Clark:2015sha,Dror:2019twh,Ramani:2020hdo,Lee:2020wfn}.
The latter two techniques are especially sensitive to the 
diffuse clumps produced by MPS enhancements considered in this work. 
These can be used to search for DM substructure in a way that is largely independent of the underlying particle properties. 

This work is organized as follows. In  Sec.~\ref{sec:pipeline} we describe the semi-analytic methods we use to predict  
clump formation times, masses, sizes, and densities, and their distributions, taking as input the MPS. We study simple power-law models for  MPS enhancements, capturing the gross features 
of many of the particle physics models described above, and consider both isocurvature and adiabatic primordial perturbations, which are associated with different transfer functions. We use the Press-Schechter ansatz to determine clump properties as a function of collapse redshift. 
In Sec.~\ref{sec:probes} we discuss the observability of microhalos produced from MPS bumps via gravitational probes or in Earth-microhalo encounters (if a non-gravitational coupling exists); there we briefly describe pulsar timing (recently studied in detail in Ref.~\cite{Lee:2020wfn} in the 
context of MPS enhancements similar to the ones considered here), and mainly 
focus on estimating caustic microlensing sensitivities and the clump-Earth encounter rates.
In Sec.~\ref{sec:model_examples} we sketch the analysis for three example models motivated by different microphysical scenarios and compare with the caustic microlensing sensitivity curves. In this section we also use the extended Press-Schechter method to estimate the subhalo distribution of the Milky Way in each of the example scenarios. In Sec.~\ref{sec:discussion} we summarize and conclude, discussing in particular the limitations of our simplified analysis. A central theme here is that we must necessarily make predictions for the DM substructure 
on very non-linear scales.  It is crucial to check these results using $N$-body simulations. Given the prevalence of small-scale structure enhancements in microphysical models of DM, together with the prospect of new detection capabilities, such studies are of very high interest.

\section{From Power Spectrum to Microhalos}
\label{sec:pipeline}
The late-time distribution of 
DM substructure can be estimated by evolving the primordial density field 
to collapse. In this paper we employ a combination of semi-analytic techniques that enable rapid exploration of clump parameter space. 
Given an initial power spectrum of DM density fluctuations, we evolve it 
to the non-linear regime using linear perturbation theory. This allows us to
estimate average properties of the collapsed objects, in particular size and density, 
and to use the Press-Schechter formalism to determine the halo distribution. Halo size and density are the key properties to which  the observational 
techniques discussed in Sec.~\ref{sec:probes} are sensitive.

The semi-analytic prescription outlined above glosses over the non-linear 
evolution of DM clumps in the presence of baryons and their potential disruption by gravitational interactions. The small-scale structures considered in this work 
are, at best, at the edge of the resolution of state-of-the art 
cosmological $N$-body simulations. We content ourselves with  simple estimates of clump disruption in our galaxy, but emphasize that continuing 
numerical studies are necessary for making more realistic projections for some astrophysical probes. (Other probes, as discussed in Sec.~\ref{sec:probes}, are primarily sensitive to clumps moving freely in galaxy clusters, and are less subject to disruptive processes.)
Below we discuss the prediction of late-time DM clump distribution 
in more detail. 

\subsection{Linear Evolution}
\label{sec:linear_evol}

If the primordial density fluctuations of DM are smaller than unity, linear 
perturbation theory can be used to evolve them forward in time. 
The evolution equations that follow from covariant stress-energy conservation are~\cite{Ma:1995ey} 
\begin{subequations}
\begin{align}
\dot \delta & = - \theta  + 3\dot \phi\\
\dot \theta & = - \frac{\dot a }{a} \theta + k^2 \psi\, ,
\end{align}
\label{eq:cdm_linear_system}%
\end{subequations}
where dots indicate derivatives with respect to conformal time, $\delta$ is the Fourier-space density contrast with comoving wavenumber $k$, $\theta=i\vec{k}\cdot \vec{v}$ is the velocity divergence, $a$ is the Friedmann-Robertson-Walker (FRW) scale factor, and $\phi$ and $\psi$ are the metric perturbations in conformal Newtonian gauge.

Different microphysical realizations of DM usually have different 
evolution equations at small scales (i.e., additional 
terms in the system above that become important for sufficiently large $k$), leading to non-trivial 
$k$ dependence for $k$ larger than some cutoff $k_c$. For example, in 
the standard WIMP case, small scales are affected by the DM 
coupling to SM radiation, leading to a non-vanishing pressure. 
This effect leads to washout of structure on scales smaller than the 
corresponding Jeans scale. In models of wave-like DM (such as axion-like particles or vector DM), $k_c$ corresponds to the comoving Compton wavelength 
of the field below which the field cannot be localized, also leading to an 
effective pressure term. In both examples, power on scales $k>k_c$ is suppressed.

Away from these small scales, the evolution of 
a DM candidate is fully specified by its equation of state (assumed to be constant during the epochs of interest), and therefore Eq.~(\ref{eq:cdm_linear_system}) is universal. 
The small-scale effects can often be modeled using the 
generalized DM framework, which modifies Eq.~(\ref{eq:cdm_linear_system}) by additional model-dependent pressure and sound speed terms~\cite{Hu:1998kj}. The modes of interest for structure formation first 
evolve through a period of radiation domination (RD, possibly 
after a period of non-standard cosmology), followed 
by matter domination (MD). 
We will consider evolution during these two phases separately. 

The formal solution of Eq.~(\ref{eq:cdm_linear_system}) during radiation domination is~\cite{Hu:1995en}
\beq
\delta = C_1 \ln a + C_2 + \int_0^\eta \left[\ln a' - \ln a\right]\frac{a'}{\dot{a}'}(k^2 \psi -3\dot \phi -3 \ddot \phi) d\eta'
\label{eq:rd_delta_sol}
\eeq
where $\eta$ is conformal time.
In Newtonian gauge, perturbations are constant prior to horizon entry, so $C_1 = 0$. The value of $C_2$ depends on whether the modes are adiabatic or isocurvature.

For adiabatic perturbations, superhorizon initial conditions give $C_2 = 3\phi/2$, 
so the second and third terms are comparable; when the mode enters the horizon 
during radiation domination, the rapidly-decaying gravitational potentials in the third term generate a logarithmically growing 
contribution. Well after horizon entry, the solution in standard cosmology is well-described by 
(see Appendix B2 of Ref.~\cite{Hu:1995en})
\beq
\delta = 
C_2 + I_1 \phi \ln \left(\frac{I_2 a}{\ahor}\right),
\label{eq:cdm_sol_rd}
\eeq
where $I_1\approx 9.11$, $I_2\approx 0.504$, $\ahor/\aeq \approx \sqrt{2}\keq/(2k)$, and $\phi$ is the 
initial (superhorizon) value of the gravitational potential.\footnote{\label{fnC2I21} While we mostly follow the notation of 
Ref.~\cite{Hu:1995en} here, note that our parameter $I_2$ in Eq.~(\ref{eq:cdm_sol_rd}) slightly differs from 
that of Ref.~\cite{Hu:1995en}, which incorporates $C_2$ into
$I_2$. It is somewhat convenient to keep them separate when discussing isocurvature and adiabatic perturbations in a unified way; when we consider the specific case of adiabatic perturbations alone, we will absorb $C_2$ into $I_2$, in which case $I_2\rightarrow 0.594$. See Footnote~\ref{fn:C2I22}.} This parametrization 
is valid for $\ahor\ll a \ll \aeq$, where $\ahor$ is the scale factor at horizon entry of mode $k$ and $\aeq$ is the scale factor at matter-radiation equality (MRE).

In models with isocurvature initial conditions, $C_2$ is not correlated with the size of gravitational potentials. This means that if $C_2 \gg \phi$, the perturbations do not grow significantly during RD. 
Note that Eq.~(\ref{eq:cdm_sol_rd}) is correct in both adiabatic and isocurvature cases. 

The part of the solution in Eq.~(\ref{eq:rd_delta_sol}) that is sourced by the gravitational potentials is 
only valid for modes that enter during standard RD. However, the form of 
the late-time solution, Eq.~(\ref{eq:cdm_sol_rd}), holds even for modes that entered during a period of non-standard cosmology, albeit with different values of  $C_2$, $I_1$ and $I_2$. These parameters are usually $k$-dependent and can be determined
 by matching the evolution of $\delta$ through a period of modified cosmology (and into RD) onto  the $\ln a$ and constant terms. For detailed examples of this matching, see  Refs.~\cite{Erickcek:2011us,Redmond:2018xty,Blinov:2019jqc}. For example, if a mode has entered the horizon during a period 
of early matter domination, then $C_2=I_1\phi = 2\phi (k/k_*)^2 /3$, $I_2 = (k_*/k)^2$ ~\cite{Erickcek:2011us,Blinov:2019jqc}. If a mode enters during a period of kination, then $C_2 = I_1\phi = 3.2\phi\sqrt{k/k_*}$, $I_2 = \sqrt{k_*/(\sqrt{2}k)}$~\cite{Redmond:2018xty}. In these expressions  $k_*$ is the comoving horizon size at the end of 
the period of modified cosmology.

Next we consider the evolution of perturbations through 
matter-radiation equality (MRE) and beyond. Well after the radiation 
contributions to gravitational potentials have decayed, the system in Eq.~(\ref{eq:cdm_linear_system}) can be cast as the Meszaros equation~\cite{Meszaros:1974tb,Hu:1995en},
\beq
\delta'' + \frac{2+3y}{2y(1+y)}\delta' - \frac{3}{2y(1+y)}\delta = 0,
\label{eq:meszaros_eq}
\eeq
where $y \equiv a/\aeq$, primes denote derivatives with respect to $y$, and we 
set the baryon density to $\Omega_b = 0$ for simplicity.
  Matching the solution of this equation in the 
$y\ll 1$ limit to Eq.~(\ref{eq:cdm_sol_rd}) gives
\beq
\delta = \frac{3}{2}\left[C_2 + I_1 \phi\ln \left(4I_2 e^{-3}\frac{\aeq}{\ahor}\right)\right]U_1(y) - \frac{4}{15}I_1 \phi U_2(y),
\label{eq:meszaros_matched_sol}
\eeq
where\footnote{The Meszaros equation~(\ref{eq:meszaros_eq}) can be solved 
for a non-zero baryon fraction $f_b$ as in Ref.~\cite{Hu:1995en}; we 
present the $f_b=0$ solutions for simplicity but use $f_b\neq 0$ in 
our numerical results.}
\beq
U_1(y) = \frac{2}{3} + y,\;\;\;\; 
U_2(y) = \frac{45}{8}\left[\left(\frac{2}{3}+y\right)\ln\frac{\sqrt{1+y}+1}{\sqrt{1+y}-1}-2\sqrt{1+y}\right]
\eeq
are the growing and decaying solutions of Eq.~(\ref{eq:meszaros_eq}). We again note that the constants $C_2$, $I_1$, and $I_2$ depend on the expansion rate of the universe when the mode entered the horizon, and can be determined by matching as described above.

In what follows we will need the autocorrelation function $\langle \delta^2 \rangle$. We write it as
\beq
\langle \delta^2 \rangle = \left(\frac{3}{2}\right)^2 \left[
\mathcal{D}_{iso}^2 \langle C_2^2 \rangle + 
\mathcal{D}_{adi}^2 I_1^2 L^2\langle \phi^2 \rangle
+ 2\mathcal{D}_{iso}\mathcal{D}_{adi} I_1 L\langle C_2 \phi \rangle\right]
\label{eq:delta_autocor}
\eeq
where
\beq
L = \ln \left(4I_2 e^{-3}\frac{\aeq}{\ahor}\right)
\label{eq:log_piece}
\eeq
is the logarithm from Eq.~(\ref{eq:meszaros_matched_sol}), 
and we defined the growth functions:
\beq
\mathcal{D}_{iso} = U_1,\;\;\; \mathcal{D}_{adi} = U_1 - \frac{8}{45 L} U_2.
\label{def:gf_def}
\eeq
At late times,
\begin{align}
    \mathcal{D}_{iso} \approx \mathcal{D}_{adi}\approx a/\aeq\;\;\;\;\;\;\;\;\;(a/\aeq\gg 1).
\end{align}

Let us consider two limits of Eq.~(\ref{eq:delta_autocor}). 
First we take the standard case of adiabatic initial conditions and $C_2 = 3\phi/2$. 
In this case, for $a/\aeq\gg 1$, it is easy to see that
\beq
\langle \delta^2 \rangle = \mathcal{D}_{adi}^2 I_1^2 L^{\prime 2} 
P_\mathcal{R},
\label{eq:adi_limit}
\eeq
where\footnote{\label{fn:C2I22}$I_2$ changes because in this particular case it is convenient to absorb $C_2$ into $I_2$. See Footnote~\ref{fnC2I21}.}
\begin{align}
    L^\prime=\ln \left(4I_2 e^{-3}\frac{\aeq}{\ahor}\right),\;\;\;\;I_2\to 0.594
    \label{eq:Lprime}
\end{align} 
and we related $\langle \phi^2\rangle$ to the curvature power spectrum
\beq
P_{\mathcal{R}}(k) = \left(\frac{3}{2}\right)^2 \langle \phi^2\rangle = 
\frac{2\pi^2}{k^3} \times A_s \left(\frac{k}{k_0}\right)^{n_s-1}.
\label{eq:curvature_ps}
\eeq
The 2018 Planck best-fit values of the spectral index and initial amplitude of curvature perturbations are $n_s=0.965\pm 0.004$ and $\ln(10^{10}A_s)=3.044\pm 0.014$  at the pivot scale $k_0=0.05~\textrm{Mpc}^{-1}$~\cite{Aghanim:2018eyx}.

In the large isocurvature limit $C_2\gg \phi$, we instead find 
\beq
\langle \delta^2 \rangle = \mathcal{D}_{iso}^2 \left(\frac{3}{2}\right)^2 
\langle C_2^2 \rangle.
\label{eq:iso_limit}
\eeq
Comparing Eqs.~(\ref{eq:adi_limit}) and~(\ref{eq:iso_limit}) we see that large isocurvature perturbations do not benefit from the radiation driving effect or the logarithmic growth during RD. 
This difference can be significant -- for example, adiabatic modes 
that enter the horizon at $T\sim \MeV$ have $I_1 L\approx 130$.
During MD, the growth is linear in scale factor 
for both types of fluctuations. This means that an isocurvature perturbation collapses significantly later than an adiabatic perturbation with the same initial amplitude.

The linear theory results described above can be used to propagate an initial power spectrum of density fluctuations (encoded in $C_2$ and $I_1$ in concrete models) until they are of $\mathcal{O}(1)$, at which point perturbation theory breaks down, signaling the onset of gravitational collapse. This, however, does not mean that microhalos begin to self-gravitate at this point. If collapse occurs during RD, the DM provides a negligible contribution to the gravitational potential, unless that overdensity is so large that the region around it is locally matter-dominated. This can occur in two distinct ways, corresponding to isocurvature and adiabatic perturbations. In the isocurvature case, the initial density fluctuation can be so large 
that local MRE is attained well before the global MRE. This was studied in Ref.~\cite{Kolb:1994fi} in a spherical collapse model. Adiabatic perturbations, on the other hand, by definition must reach collapse dynamically. Ref.~\cite{Blanco:2019eij} 
studied the case where this occurs during the logarithmic growth of a 
perturbation (originally enhanced through a period of EMD). 
In this example, the perturbation becomes non-linear well before it begins to self-gravitate and virialize (when local MRE is attained). 

In what follows, we focus on fluctuations that collapse only after global MRE, enabling the use of standard results on spherical collapse and avoiding the difficulties of collapse without virialization. While not fully general, this restriction 
still captures many of the scenarios introduced in Sec.~\ref{sec:intro}. Moreover, we expect that the 
gravitational probes discussed later are more sensitive to 
objects that collapse at or after MRE, given the range of mass scales that they can probe.\footnote{Other probes, such 
as indirect detection, may be significantly enhanced in objects that form earlier~\cite{Blanco:2019eij}.}

\subsection{Collapse and the Formation of Microhalos}
\label{sec:collapse}

The Press-Schechter (PS) formalism enables a semi-analytic understanding of structure 
formation from the evolution of the linear density contrast discussed above~\cite{Press:1973iz}. 
PS postulates that the fraction of matter in collapsed objects of size 
$R$ at a given time is related to the cumulative probability distribution 
to find $\delta_R > \delta_c$, where $\delta_R$ is the density contrast 
smoothed over scales $R$ and $\delta_c$ is the critical linear density contrast, i.e., the fractional overdensity that would collapse in a non-linear treatment.  For spherical collapse during MD, $\delta_c \approx 1.686$~\cite{Galform:2010}. The second key assumption in PS is that the 
probability distribution of $\delta_R$ is Gaussian and therefore fully
specified by its variance,
\beq
\sigma^2(z, R) = \int \frac{d^3 k}{(2\pi)^3} \langle \delta(a,k)^2 \rangle W(kR)^2,
\label{eq:density_variance_def}
\eeq
where $W$ is a smoothing kernel and we use redshift $z= 1/a - 1$ and the scale factor $a$ interchangeably as independent variables. The comoving smoothing scale $R$ can be translated to a mass scale $M \propto \bar\rho_0 R^3$ ($\bar\rho_0$ is the average DM density today), with 
the precise relationship between $M$ and $R$ depending on the choice of $W$ (see Appendix~\ref{sec:wf}).

The variance $\sigma^2$ encodes the evolution of structure as a
function of time. A useful quantity that can be derived from it 
is the critical mass $M_*(z)$ such that 
\beq
\sigma(z,M_*) = \delta_c.
\label{eq:mstar_def}
\eeq
At a given redshift, $M_*$ gives the typical mass of the largest structures that collapse at this time.\footnote{Here ``typical'' means $1\sigma$ 
fluctuations at redshift $z$. Objects of a fixed mass $M_*(z)$ can form 
earlier than $z$, but they must arise from rarer upward fluctuations in the density contrast.}
In all microphysical models, $M_*$ is a monotonically increasing function of $1/(1 + z)$, indicating the hierarchical assembly of smaller clumps into larger and 
larger structures. The density fluctuation variance $\sigma$ and the characteristic 
mass $M_*$ can be used to model the distribution of microhalos as a function 
of redshift within the PS framework and its extensions. We will do this in Sec.~\ref{sec:shmf}, but for now we briefly describe the basic properties of
the earliest-forming microhalos.

In the spherical collapse model, the average density of matter in a newly-formed object depends only on the collapse redshift $z_c$~\cite{Peebles:1980,Cole:1995ep,Galform:2010}:
\begin{align}
    \rho(z_c) \approx 178\bar\rho(z_c)\approx 3500{\rm~ GeV/cm}^3\,\left(\frac{1+z_c}{250}\right)^3,
\label{eq:rhozc}
\end{align}
where $\bar\rho(z_c)$ is the background density at the time of collapse. 
Then a characteristic radius of clumps forming at redshift $z_c$ can be assigned as
\begin{align}
    R_*(z_c) =\left(\frac{3 M_*(z_c)}{4\pi \rho(z_c)}\right)^{1/3}.
    \label{eq:rstar_def}
\end{align}
We see that objects that form earlier are denser and more compact. In Sec.~\ref{sec:disruption} we will argue that earlier-forming microhalos are 
more likely to survive tidal disruption by stars, other clumps, and the galactic 
halo.

After a clump forms, its early-time evolution depends on whether the bump in the power spectrum is a localized spike or a broad enhancement~\cite{Delos:2017thv,Delos:2018ueo}. For spiked features, a narrow range of scales goes nonlinear at approximately the same time, while all other scales must evolve for a much longer period before crossing the collapse threshold. In this case, the clumps that form first evolve for a long time in relative isolation, and they have featureless profiles with peaked inner densities scaling as $\rho\sim r^{-3/2}$. In the case of broad enhancements, larger and larger clumps are continually forming, coalescing out of their predecessors. Thus the enhanced population of microhalos develops with a substructure of smaller, older clumps, which is expected to survive at least an order of magnitude in redshift after formation~\cite{Delos:2017thv,Delos:2018ueo}. The average density profiles in this case are modified by accretion, trending toward the Navarro-Frenk-White (NFW)~\cite{Navarro:1995iw} profile with $\rho\sim r^{-1}$. 
In the examples of Sec.~\ref{sec:power_law_examples} and~\ref{sec:model_examples} we will consider both narrow and broad enhancements to the MPS. For simplicity, however, in both cases we will model the density profile as NFW. For a narrow spike, this choice underestimates how compact the resulting objects are, and therefore 
overestimates the disruption probability. This choice is thus conservative in estimating the late time abundance of these objects. 

The NFW profile is given by
\beq
\rho(r) = \frac{4\rho_s}{(r/r_s)(1+r/r_s)^2},
\eeq
where $\rho_s$ and $r_s$ are scale mass and scale radius.
The mass within a given radius $r$ is then
\beq
M(r) = 16\pi \rho_s r_s^3 f(r/r_s),
\eeq
where 
\beq
f(c) \equiv \ln(c+1) - \frac{c}{c+1}.
\eeq
The scale mass is the mass within the scale radius:
\beq
M_s \equiv M(r_s) = 16\pi \rho_s r_s^3 f(1).
\label{eq:scale_mass_def}
\eeq
We can approximately relate the collapse mass $M_*$ and radius $R_*$, given 
by Eqs.~(\ref{eq:mstar_def}) and~(\ref{eq:rstar_def}), to $r_s$, $M_s$, and $\rho_s$. Recall that $R_*$ is defined as the radius within which 
the density is $178\bar\rho(z_c)$, where $\bar\rho(z_c)$ is the 
background density at collapse. This means that $M_*$ and $R_*$ 
are close to the virial quantities $M_{200}$ and $R_{200}$, defined for halos with an average density 200 times the background density, evaluated at collapse. 
Simulations of $\Lambda$CDM halos suggest that the concentration parameter $c_{200}=R_{200}/r_s$
shortly after formation is about $2$ for Earth-mass halos in $\Lambda$CDM~\cite{Sanchez-Conde:2013yxa} (one needs to 
take the Earth mass $c_{200} \approx 60$ and redshift it back to $z \approx 30$ where these halos were measured in the first place), 
implying that $M_*$ and $R_*$ are similar to the NFW scale quantities. 
Let $c_*$ be the concentration parameter at formation, i.e.,
\beq
c_* = R_*/r_s.
\eeq
Let us also assume that $R_{200}(z_c) = R_*$, neglecting the difference between $200$ and $178$. 
This allows us to find $M_s$ and $\rho_s$ from $c_*$, $R_*$ and $M_*$:
\beq
M_s = \frac{f(1)}{f(c_*)} M_*.
\eeq
The scale density is then obtained from Eq.~(\ref{eq:scale_mass_def}). 
As mentioned above, halos of different masses can form at around the same redshift. Thus a more realistic description of the microhalo population would account for scatter in the concentration parameter $c_*$~\cite{Cooray:2002dia}. This scatter reflects both the ``rarity'' of a collapsing density peak (or, equivalently, the spread in formation redshifts of 
halos with similar masses)~\cite{Wechsler:2001cs} and the subsequent assembly history. We will focus on a single representative value of $c_*$ for simplicity.

\subsection{Scaling Behaviors}
\label{sec:scalings}

Here we derive some simple scaling laws that allow estimation of the properties of the largest clumps forming from an enhancement in the power spectrum. 
We modify the primordial power spectrum in Eq.~(\ref{eq:curvature_ps}) by introducing a ``bump'' function $B$ that enhances the power spectrum over some range of scales, such that $P_{\mathcal{R}}(k)\to P_{\mathcal{R}}(k)B(k)$.
The resulting variance in the density contrast on a given comoving scale $ \hat k$ at scale factor $a$ is:
\begin{align}
\sigma^2(a,R=1/\hat k) &= \frac{1}{2\pi^2}\int dk \, k^2 W(k/\hat k)^2|D(a,k)|^2 P_{\mathcal{R}}(k)B(k)\nonumber\\
&\sim A_s D(a,\hat k)^2 \left(\frac{\hat k}{k_0}\right)^{n_s-1} \left(\frac{\hat k}{\bar k}\right)^{n_1}.
\end{align}
Here $W$ is a window function that isolates scales of order $\hat k$ in some way. In the second line we assume that $B\sim (k/\bar k)^{n_1}$ characterizes the nontrivial behavior of $B$ at the largest enhanced length scales, where $\bar k$ is the smallest value of $k$ with any enhancement, and we focus on those scales since they will determine the largest substructures. The relevant scale-dependent growth function $ D(a,k)$ depends on whether the bump is isocurvature or adiabatic:
\begin{align}
D(a,\hat k) \sim \left(\frac{a}{\aeq}\right)  L^\alpha \;,~~~~~~
L \sim 10 \ln(\aeq/\ahor)\sim 10 \ln( \hat k/\keq),
\end{align}
where the scale factor and $k$ dependence follows from  Eqs.~(\ref{eq:adi_limit}) and~(\ref{eq:iso_limit}), and the exponent $\alpha$ is 1 (0) for adiabatic (isocurvature) perturbations.\footnote{The factor $\langle \delta^2\rangle \propto |D(a,k)|^2 P_\mathcal{R}$ is sometimes written as $|T(k)D(a)|^2 P_{\Lambda\mathrm{CDM}}$, 
where $T(k)$ is a transfer function, $D(a) = a$, and $P_{\Lambda\mathrm{CDM}} \sim (k/\keq)^4 P_\mathcal{R}$~(see, e.g., the textbooks~\cite{Lyth:2009zz,Galform:2010}). The conventional normalization of $P_{\Lambda\mathrm{CDM}}$ is chosen such that $T(k) \approx 1$ (or $L^\alpha(\keq/k)^2$) for modes that enter the horizon after (or before) matter-radiation equality.
These conventions and normalizations are convenient for studying wavenumbers in the vicinity of $\keq$, but are somewhat encumbering in our case of interest with $k\gg \keq$, where the factors of $k/\keq$ 
always cancel between $T(k)^2$ and $P_{\Lambda\mathrm{CDM}}$.}

\begin{figure}
    \centering
    \includegraphics[width=0.47\textwidth]{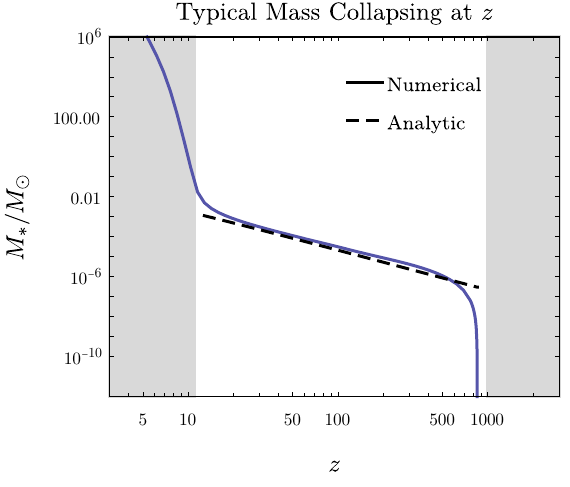}
    \caption{Comparison of the critical mass $M_*(z)$ derived by numerically solving Eq.~(\ref{eq:mstar_def}) (solid line) with the analytic estimate in Eq.~(\ref{eq:Mstarz}) (dashed line). 
    This example corresponds to an adiabatic power-law enhancement of the MPS 
    defined in Eq.~(\ref{eq:powerlaw_bump_def}) with $n_1=n_2=3$, $k_p/\keq = 10^8$ and 
    $\mathcal{E} = 10^4$. In the shaded grey regions the estimate of Eq.~(\ref{eq:Mstarz}) is not valid, since the power spectrum changes its 
    logarithmic slope; the low-$z$ boundary is found by setting 
    the factor in parenthesis in Eq.~(\ref{eq:Mstarz}) to 1, while the high-$z$ 
    boundary corresponds to collapse of the peak scale $k_p$.}
    \label{fig:mstar_example}
\end{figure}

Then the collapse redshift $z_c(\hat k)$ is determined by setting the critical linear density contrast equal to the variance: 
\begin{align}
\delta_c  \sim \sqrt{A_s}\left(\frac{\hat k}{\bar k}\right)^{n_1/2}\left(\frac{\hat k}{\keq}\right)^{(n_s-1)/2}  \left(\frac{\zeq}{z_c(\hat k)}\right) L^\alpha .
\end{align}
Turning the perspective around, we can find the mode $k_*(z)$  that characterizes scales collapsing at $z$. If we associate a mass with $k_*$ as $M_*\sim \bar\rho(a) (k_*/a)^{-3}\sim \bar\rho_0/k_*^3$ and set $n_s\approx 1$, we have 
\begin{align}
\delta_c  \sim \sqrt{A_s}\left(\frac{\overline M}{ M_*}\right)^{n_1/6}  \left(\frac{\zeq}{z}\right) L^\alpha
\end{align}
where $\overline{M} \sim \bar\rho_0/\bar k^3$ and
\begin{align}
M_*(z) \sim \overline{M}\left(\frac{ \zeq L^\alpha \sqrt{A_s}}{z \delta_c}\right)^{6/n_1}.
\label{eq:Mstarz}
\end{align}

In Fig.~\ref{fig:mstar_example} we compare the analytic estimate in Eq.~(\ref{eq:Mstarz}) (dashed line) with the critical mass $M_*$ derived by solving Eq.~(\ref{eq:mstar_def}) numerically (solid line) for a particular choice of $B$ defined below in Eq.~(\ref{eq:powerlaw_bump_def}). We see that  Eq.~(\ref{eq:Mstarz}) provides an excellent approximation to the critical mass over nearly two orders of magnitude in $z$. 
In the shaded grey regions the estimate of Eq.~(\ref{eq:Mstarz}) is not valid, since the power spectrum changes its logarithmic slope; the low-$z$ boundary is found by setting the factor in parenthesis in Eq.~(\ref{eq:Mstarz}) to 1, while the high-$z$ boundary corresponds to collapse of the peak scale $k_p$ where the power law growth with $n_1$ ends (see Eq.~(\ref{eq:powerlaw_bump_def})).

\subsection{Results for Power Law Peaks}
\label{sec:power_law_examples}

We use simplified models to parametrize  bumps in the power spectrum. We add power-law enhancements over the primordial $\Lambda$CDM baseline in Eq.~(\ref{eq:curvature_ps}), specifying a wavenumber $k_p$ which determines the location of the peak of the bump in the power spectrum and a numerical factor $\mathcal{E}$ which gives the size of the enhancement over the value for $\Lambda$CDM at the same value of $k_p$. We assume that the rise and fall in the power spectrum on either side of peak are power laws in $k$, which we allow to have different exponents $n_1$ and $n_2$. The final form of the peak enhancement can be written as
\beq
B(k) = \mathcal{E} \times
\begin{cases}
(k/k_p)^{n_1} & k < k_p \\
(k_p/k)^{n_2} & k \geq k_p \, ,
\end{cases}
\label{eq:powerlaw_bump_def}
\eeq
such that the full power spectrum is given by $P_{\mathcal{R}}(k)$ 
for $k < k_p/\mathcal{E}^{1/n_1}$ and $B(k)P_{\mathcal{R}}(k)$ otherwise.
This simple expression captures the main features of a broad range of well-motivated models. For instance, an early epoch of kination predicts $n_1 = 1$~\cite{Redmond:2018xty}, while early matter domination gives $n_1=4$~\cite{Erickcek:2011us}. The inflationary production of massive scalar spectator fields can generate 
$0 \leq n_1 \leq 3$~\cite{Chung:2004nh}, while ultralight dark vector DM gives $n_1=3$~\cite{Graham:2015rva}.
Meanwhile, altering the falling slope on the right hand side of the peak approximates a number of physical cutoff mechanisms. A very rapidly falling peak, which we can model as e.g. $n_2=\infty$, can correspond to a Jeans cutoff, related to the wave nature of a DM candidate~\cite{Blinov:2019jqc} or its coupling to the radiation bath~\cite{Erickcek:2011us}. The other extreme $n_2=0$, where the rise is followed by a plateau, can be found in models where a modified expansion period is relatively short and preceded by radiation domination~\cite{cannibalII}. (In such cases the plateau should reflect logarithmic growth; this is a minor effect that is not captured by the simplified model.) An intermediate regime, where the fall is less steep than the rise, can be found in models of dark photon DM produced by inflationary fluctuations, where $n_2=1$~\cite{Graham:2015rva} or EMD models with cannibal interactions, where $n_2=2$ \cite{Erickcek:2020wzd}. We discuss these three scenarios in further detail in Sec.~\ref{sec:model_examples}.

This simplified model lets us study the statistical properties of dark matter clumps as a function of the peak location, peak height, and the power laws on the rising and falling sides of the peak, for both adiabatic and isocurvature perturbations. These are the most important data affecting final distributions and observables. We take Eq.~(\ref{eq:powerlaw_bump_def}) to parametrize an ``effective'' primordial power spectrum, and then use the standard growth functions derived in Sec.~\ref{sec:linear_evol} to evolve them forward in time. The effect of any non-standard cosmological evolution is absorbed into $B(k)$, even though it can occur well after inflation. More explicitly, in Eq.~(\ref{eq:density_variance_def}) we use
\beq
 \langle \delta(a,k)^2 \rangle
 = \begin{cases}
 \mathcal{D}_{adi}(a)^2 I_1^2 L^{\prime}(k)^2 
B(k)P_{\mathcal{R}}(k) & \text{(adiabatic)}\\
\mathcal{D}_{iso}(a)^2 \left[B(k) -1 \right]P_{\mathcal{R}}(k) + \mathcal{D}_{adi}(a)^2 I_1^2 L^{\prime}(k)^2 
P_{\mathcal{R}}(k)& \text{(isocurvature)}
 \end{cases},
\eeq
where the curvature power spectrum $P_\mathcal{R}$ is given in Eq.~(\ref{eq:curvature_ps}) and the various growth factors are for the standard cosmology. For reference, growth functions $\mathcal{D}$ are defined in Eq.~(\ref{def:gf_def}), $I_1$ is defined below Eq.~(\ref{eq:cdm_sol_rd}), and $L'$ is given in Eq.~(\ref{eq:Lprime}). Note that when $B(k)$ represents an isocurvature enhancement,  we assume that the MPS is still the standard adiabatic one at large scales; this is enforced in the second line by the presence of $B(k) - 1$, which vanishes for $k < k_p/\mathcal{E}^{1/n_1}$.
We now consider varying each of the MPS parameters in Eq.~(\ref{eq:powerlaw_bump_def}) in turn, measuring the sensitivity of the properties of the resulting microhalos.

\begin{figure}[t!]
\begin{center}
\includegraphics[width=0.47\textwidth]{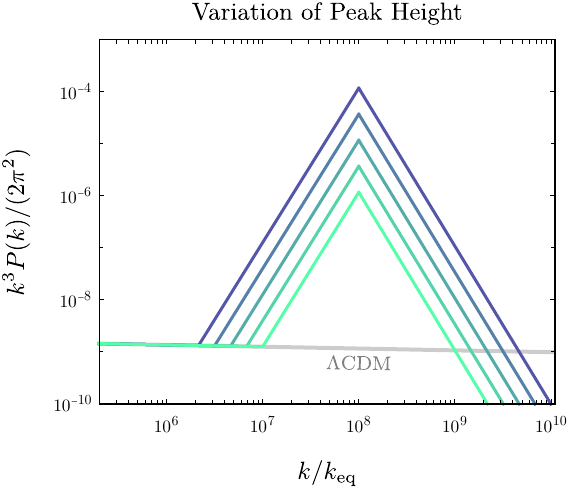}\;\;
\includegraphics[width=0.47\textwidth]{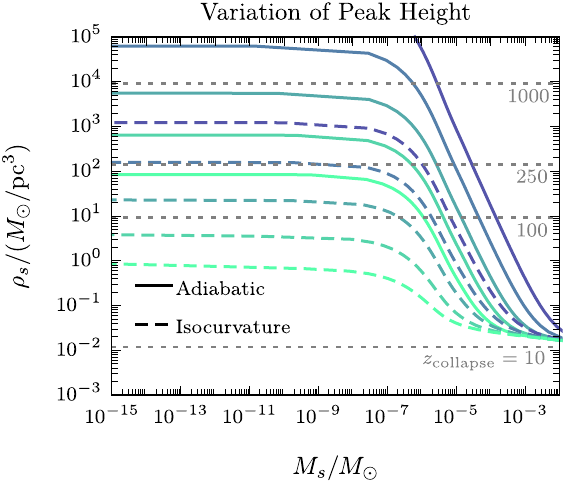}\\
\includegraphics[width=0.47\textwidth]{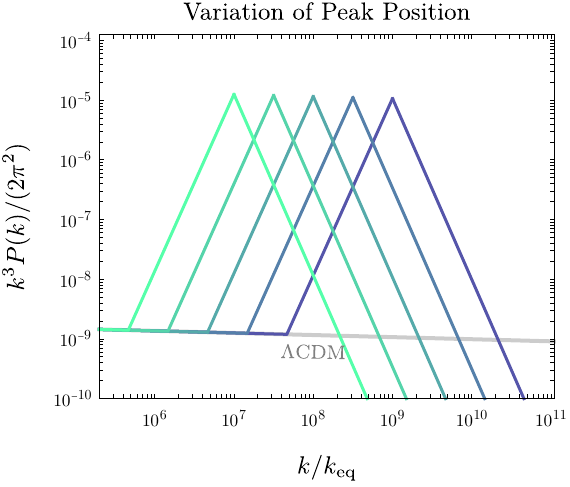}\;\;
\includegraphics[width=0.47\textwidth]{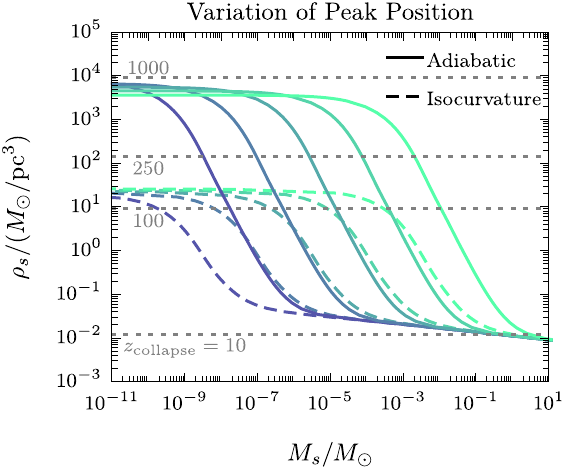}\\
\caption{Power spectra and the resulting microhalo properties for different peak amplitudes (top row) and positions (bottom row). Upper left:  The dimensionless power spectrum for different peak heights for $\mathcal{E} = 10^{4 +i/2}$ with $i \in \{-2,-1,0,1,2\}$ (corresponding to the gradation of lighter to darker lines) in  Eq.~\ref{eq:powerlaw_bump_def}; peak position $k_p/\keq = 10^8$ and slopes $n_1=n_2 = 3$ are held fixed. Upper right: Minihalo scale density as a function of scale mass, $\rho_s(M_s)$, for these peak height variations, showing both adiabatic (solid) and isocurvature (dashed) perturbations. Dashed grey lines indicate the collapse epoch $z_c$.
Lower left: The power spectrum for different peak positions $k_p/ \keq=10^{8+i/2}$ with $i \in \{-2,-1,0,1,2\}$ (corresponding to the gradation of lighter to darker lines); the enhancement is 
held fixed at $\mathcal{E}=10^4$. Lower right: As for upper right, showing varying peak position. Note the scales of the axes in the right-hand plots differ.}
\label{fig:peakstats_1}
\end{center}
\end{figure}

In Fig.~\ref{fig:peakstats_1} we vary the height of the peak in the top line as $\mathcal{E}= 10^{4 +i/2}$ for $i \in \{-2,-1,0,1,2\}$. The peak location is kept fixed at $k_p/ \keq=10^8$ and the slopes to $n_1=n_2=3$. These enhancements to the power spectrum are shown in the upper-left panel. The upper right panel shows the properties of the collapsed clumps as in the $M_s$-$\rho_s$ plane. The scale mass $M_s$ and scale density $\rho_s$ can be related to the critical mass $M_*$ and radius $R_*$ assuming a value for the concentration parameter at formation, as described in Sec.~\ref{sec:collapse}.

Each point on the lines in the right-hand panels of Fig.~\ref{fig:peakstats_1} corresponds to a specific value of the collapse epoch $z_c$ for the clumps of that mass, since $z_c$ relates the scale mass to the scale radius (and hence scale density).  The horizontal grey-dashed lines correspond to collapse epochs of $z_c=10,\,100,\,250$ and $1000$ from bottom to top, respectively. As we discuss in Sec.~\ref{sec:disruption} clumps which form at redshifts prior to $z=250$ are unlikely to be severely disrupted in our galaxy. 

\begin{figure}[t!]
\begin{center}
\includegraphics[width=0.47\textwidth]{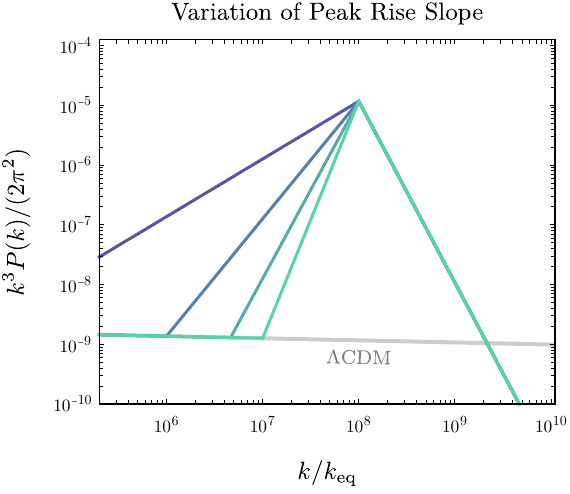}\;\;
\includegraphics[width=0.47\textwidth]{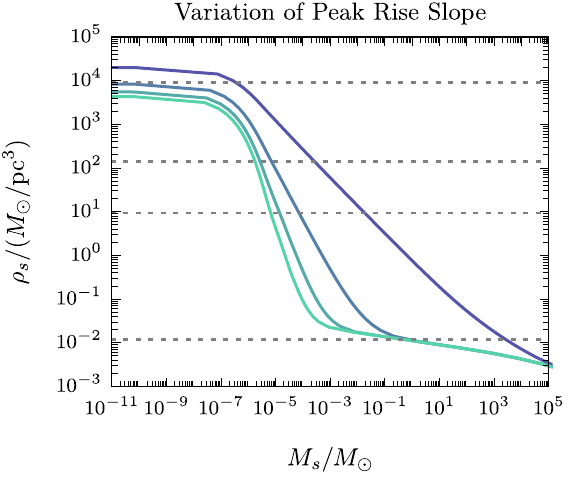}\\
\includegraphics[width=0.47\textwidth]{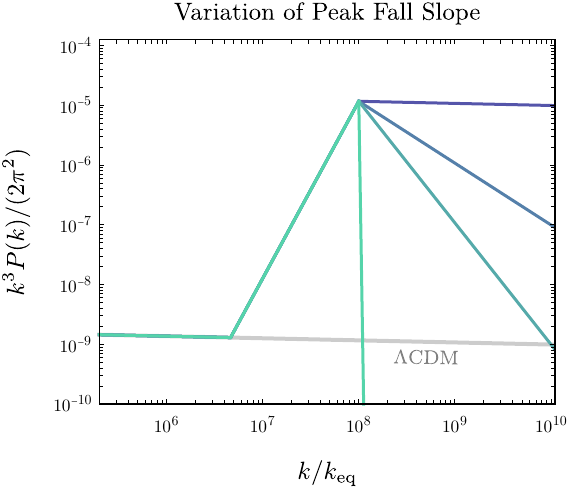}\;\;
\includegraphics[width=0.47\textwidth]{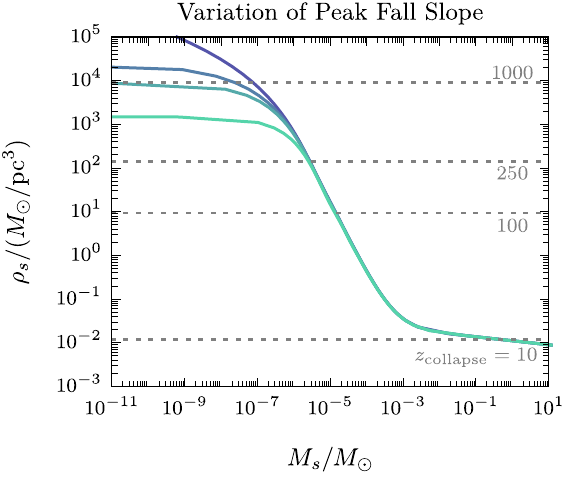}\\
\caption{ Power spectra and the resulting microhalo properties for different rise (top row) and fall slopes (bottom row). Upper left: Rise slope variations for $n_1 \in \{1,2,3,4\}$ (corresponding to the gradation of darker to lighter lines) in Eq.~\ref{eq:powerlaw_bump_def}, with fixed $n_2=3$, $k_p/\keq = 10^8$ and $\mathcal{E}=10^4$. Upper right: Minihalo scale density as a function of scale mass, $\rho_s(M_s)$ for the different rise slopes. Dashed grey lines indicate the collapse epoch $z_c$.
Lower left: Power spectra for different peak fall slopes $n_2\in \{0,1,2,\infty\}$ (corresponding to the gradation of darker to lighter lines), with fixed $n_1=3$, $k_p/\keq = 10^8$ and $\mathcal{E}=10^4$. Lower right: As for upper right, showing varying peak fall slopes. Note the scales of the axes in the right-hand plots differ.}
\label{fig:peakstats2}
\end{center}
\end{figure}

We present results for both adiabatic (solid) and isocurvature (dashed) perturbations. Increasing the enhancement causes modes of a given $k$ to go nonlinear earlier. Fluctuations on physical scales to which those modes contribute significantly thus collapse earlier, increasing the density of the resulting clump. For adiabatic perturbations, enhancements greater than $\mathcal{E}=10^3$ lead to the formation of clumps which are likely to survive in the environment of the Milky Way to the present day. Isocurvature perturbations require an enhancement factor of $\mathcal{E}\sim10^{4.5}$ for the same result. Since isocurvature fluctuations do not benefit from logarithmic growth or radiation driving during RD (see the discussion in Sec.~\ref{sec:linear_evol}) they require a larger enhancement in order to collapse at the same time as an equivalent adiabatic mode.

The lower panels of Fig.~\ref{fig:peakstats_1} vary the peak position as $k_p/ \keq=10^{8+i/2}$ for $i \in \{-2,-1,0,1,2\}$, keeping the enhancement factor fixed to $\mathcal{E}=10^4$. Shifting the peak toward lower $k$ means that larger scales become non-linear earlier on, leading to an increase in the density for a given mass. This also leads to an increase in the maximum clump mass that receives a density enhancement over $\Lambda$CDM.

In Fig.~\ref{fig:peakstats2} we present results for varying the slope on the left-hand side of the peak (at lower values of $k$, upper plots), and the right-hand side of the peak (at higher values of $k$, lower plots). We keep the peak location and enhancement fixed at $k/\keq=10^8$ and $\mathcal{E}=10^4$ in these plots. In the upper panels we take $n_1\in\{2,3,4,5\}$ and in the lower panels $n_2\in\{0,1,2,\infty\}$, with the other exponent fixed to $3$. 

Fattening the low-$k$ tail causes longer wavelength modes to collapse earlier if we hold $k_p$ and $\mathcal{E}$ fixed. It increases the largest length scales associated with the MPS enhancement, and thus the mass of the largest objects with density enhancement over the ordinary $\Lambda$CDM scenario.  On the high-$k$ side of the peak, broadening the cutoff slope with $k_p$ and $\mathcal{E}$ fixed pushes the collapse epoch $z_c$ of the smallest structures to higher redshifts, resulting in denser objects.

We do not show results for isocurvature enhancements in Fig.~\ref{fig:peakstats2}. These would follow the same general trends of Fig.~\ref{fig:peakstats_1}: compared to adiabatic enhancements, isocurvature bumps would result in smaller characteristic densities and masses due to the absence of the radiation driving effect at early times.

\subsection{Microhalo Substructure, Evolution, and Disruption}
\label{sec:disruption}
The survival of early-forming microhalos until late times is paramount to 
their possible observation using the methods described in Sec.~\ref{sec:probes}. 
Gravitational interactions with other clumps, the galactic halo, and baryonic objects can transfer energy to a clump, changing its internal structure, stripping away some of its matter or disrupting it completely. The early-time hierarchical assembly of smaller structures into larger ones tends to flatten the core density profiles~\cite{Delos:2017thv,Delos:2018ueo}; this effect is more pronounced in broad MPS enhancements and motivates the NFW profile we use to model these objects, as discussed in Sec.~\ref{sec:collapse}. At later times, microhalos accreted onto the host galactic halo can be stripped of a large fraction of their mass. This mass loss occurs beyond a tidal radius $R_t$ from the microhalo center; $R_t$ is estimated as the radius beyond which the tidal force of the host halo is larger than the microhalo's self-gravity~\cite{vandenBosch:2017ynq}:
\beq
R_t(R) \approx R\left[\frac{M(R_t)}{M_h(R)}\right]^{1/3},
\label{eq:tidal_radius_estimate}
\eeq
where $R$ is the radius of the microhalo orbit around the host, and $M_h(R)$ is the 
host mass interior to $R$.
Thus $R_t/r_s \sim (\rho_s/\rho_h)^{1/3}$, where $\rho_h$ is the 
average density of the host, so $R_t/r_s \gg 1$ for early-forming microhalos within galaxy-sized hosts. This suggests that these tidal effects tend to modify the outer structure of the microhalos, but do not affect survival of the microhalo ``core'' 
with $r\lesssim r_s$. As an explicit example we can estimate the tidal radius for microhalos orbiting the MW at $R\approx R_\oplus\approx 8.1$ kpc with $M_h (R) \approx 10^{11} M_\odot$~\cite{2020MNRAS.494.4291C}. This is relevant for the pulsar timing probe of microhalos~\cite{Lee:2020wfn}. We find that $R_t/r_s > 1$ as long as the microhalos formed earlier than $z\sim 10$. In all examples we consider in this work, the MPS enhancements lead to formation of microhalos at much earlier times, and 
therefore we do not expect tidal disruption by the host halo to significantly alter these objects at late times.\footnote{This conclusion might not hold for all microhalos, such as those on highly radial orbits that take them close to the center of the host.}
Similar arguments hold for diffuse populations of microhalos (i.e., microhalos not bound to any galaxy) in galaxy clusters; in this case, it is clear from Eq.~(\ref{eq:tidal_radius_estimate}) that tidal disruption by the host is even less relevant than in our galaxy.

Stellar encounters at late times are a particularly efficient source of clump disruption, capable of destroying microhalos that formed even at $z\gg 10$.
In the impulse approximation, the specific impulse transferred to the clump constituents from an encounter with a star with mass $M_\star$ and impact parameter $b$ is of order
\begin{align}
\Delta v \sim \left(\frac{G M_\star r_{vir}}{b^3}\right)\times\left(\frac{b}{ v_{rel}}\right)
\end{align}
where the first factor is the characteristic tidal force and the second is the encounter timescale. Therefore the energy transfer relative to the binding energy of the clump is of order~\cite{Goerdt:2006hp,Schneider:2010jr}
\begin{align}
\frac{\Delta E}{E_b}\sim\frac{v_{vir}\Delta v}{v_{vir}^2} \sim  \frac{G M_* r_{vir}}{b^2 v_{rel} v_{vir}}.
\end{align}
The ratio $r_{vir}/v_{vir} \propto \sqrt{\rho}$, where $\rho$ is the clump mass density, so the density is the only microhalo property on which the relative energy transfer depends in this approximation. Setting $\Delta E/E_b=1$ defines a critical impact parameter~\cite{Goerdt:2006hp,Schneider:2010jr}
\begin{align}
b_c^2 = \frac{\sqrt{G\rho}M_\star}{v_{rel}}.
\end{align} 
Encounters with larger $b$ transfer less energy, but are also more probable. The probability to transfer a total energy of order $E_b$ over many subcritical encounters is comparable to the probability of a single critical encounter~\cite{Goerdt:2006hp,Schneider:2010jr}, so the total probability of transferring $\Delta E\sim E_b$ is roughly double the probability of a single critical encounter.

On the other hand, critical encounters with $b<b_c$ may not completely destroy the clump, because the transferred energy may not be efficiently redistributed before the outer layers are stripped away~\cite{vandenBosch:2017ynq}. In this case, a dense remnant core remains. Nonetheless, the probability of a critical encounter is a reasonable estimate for the survival rate of larger clumps. In particular, if the clumps contain substructure, the smaller, denser constituents may be spilled out by disruptive encounters, with their own structure preserved.

We define the total disruption probability $P$ to be twice the critical encounter probability~\cite{Goerdt:2006hp}
\begin{align}
P = 2n \pi b_c^2 S
\label{eq:disruption_propbability}
\end{align}
 where $n$ is the number of galactic disc crossings and $S$ is the orbit-averaged stellar column mass density along the clump orbit.\footnote{A more sophisticated treatment of the stellar distribution yields a similar disruption probability~\cite{Dokuchaev:2017psd}.} For clumps in the Milky Way, $n\sim 100$ and $S\sim 140 M_{\odot}/{\rm pc}^2$. Since $b_c$ depends only on the clump density, which in turn depends only on the redshift of collapse, we can estimate~\cite{Tinyakov:2015cgg}
\begin{align}
P \approx \frac{n}{100}\left(\frac{250}{z_c}\right)^{3/2}.
\end{align}
Therefore, we arrive at a qualitative estimate that clumps forming before $z\approx 250$ typically survive, while clumps forming later are typically strongly disrupted. A similar calculation of the survival probability has been validated in $N$-body simulations of a microhalo traversing a dense field of stars~\cite{Schneider:2010jr}.

While the pulsar timing techniques described in Sec.~\ref{sec:probes} probe the MW microhalo distribution, caustic microlensing is mainly sensitive to the diffuse microhalo population in galaxy cluster lenses. These clumps can be disrupted by the diffuse stars in the cluster. We can use Eq.~(\ref{eq:disruption_propbability}) to estimate the disruption probability. Taking some representative cluster parameters $v_{rel} \sim 1000\;\mathrm{km/s}$, $S \sim 10^7 M_\odot/\mathrm{kpc}^2$, and $n\approx 10$ we find that disruption is significant only for objects that formed after $z_c \lesssim 6$.\footnote{See, e.g., Ref.~\cite{Pillepich:2017fcc} for numerical studies of the diffuse stellar population in clusters and Ref.~\cite{Kelly:2017fps} for relevant parameters of a cluster where caustic microlensing was observed.} We therefore expect clumps originating from MPS enhancements to survive in the diffuse cluster environment.

We overlay contours of $z_c$ in the plots, both to give a sense for the pace of structure formation once the bump in the power spectrum starts to go nonlinear, and to estimate the sensitivity of the clump population and the observables to the variations in the survival cutoff on $z_c$.

\subsection{Subhalo Distribution}
\label{sec:shmf}

In this subsection we estimate the microhalo distribution inside  galaxies. 
The density fluctuation variance in Eq.~(\ref{eq:density_variance_def}) 
can be used to construct a global (Universe-averaged) distribution 
function within the Press-Schechter formalism~\cite{Press:1973iz}.
In this framework, the fraction $df$ of matter in objects of mass in the range $[M,M+dM]$ is
\beq
\frac{df}{dM}(M,z) = 
\sqrt{\frac{2}{\pi}} \frac{\delta_c}{M\sigma} \left|\frac{d\ln \sigma}{d\ln M}\right| \exp\left(-\frac{\delta_c^2}{2\sigma^2}\right).
\label{eq:PS_fraction}
\eeq
The halo function is the comoving number density of collapsed objects in this mass range:
\beq
\frac{dn}{dM}(M,z) = \frac{\bar\rho_0}{M} \frac{df}{dM},
\label{eq:PS_hmf}
\eeq
where $\bar\rho_0$ is the DM density today. Generalizations of the PS ansatz exist (such as Ref.~\cite{Sheth:1999mn,Cooray:2002dia}) that provide a 
better fit to $N$-body simulations. However, these 
simulations have been mostly performed for standard power spectra. (Exceptions include, e.g., Refs.~\cite{Delos:2017thv,Delos:2018ueo,Delos:2019mxl},  which study power spectrum enhancements from a modified cosmology. Post-inflationary axion DM models are another example widely studied in simulation; for a recent analysis, see~\cite{Xiao:2021nkb}.)
 It is therefore important to validate the predictions of the Press-Schechter ansatz and its extensions for other primordial power spectra.

Note that $M$ in Eq.~(\ref{eq:PS_fraction}) is the mass 
of recently-formed objects, i.e., those overdensities that are just collapsing at $z$. Evaluating the mass function at $z=0$ as a proxy for the distribution of microhalos in the MW (or other galaxies or clusters relevant for caustic microlensing) misses two important effects: disruption of clumps, as described in the previous section, and the fact that the objects forming now are assembled from smaller objects; disruption may ``spill'' these constituents such that these ``sub-microhalos'' are the relevant clumps at late times. A coarse model that side-steps these issues is to 
simply evaluate the distribution function at an earlier time, e.g., $z\sim 250$,
such that microhalos formed during this era are not likely to be disrupted. We will take this approach.

The PS ansatz, Eq.~(\ref{eq:PS_fraction}),  has another shortcoming in the present context -- it does not capture the fact that microhalos that end up in galaxies are necessarily embedded in larger-scale overdensities that form galaxies in the first place. As a result, the large-scale overdensity effectively lowers the collapse threshold for microhalos at early times (while the galaxy-scale overdensity is still in the linear regime), resulting in a slight enhancement of the microhalo distribution function inside of galaxies, compared to the global average. This effect is 
captured by the extended Press-Schechter (EPS), or excursion set, formalism~\cite{Bond:1990iw,Lacey:1993iv,Zentner:2006vw}, in which the 
\emph{conditional} probability distribution of microhalos of mass $M_2$ identified at $z_2$ to end up in large-scale overdensities of mass $M_1$ and (later) redshift $z_1$ is given by
\beq
\frac{df}{dM_2}(M_2,z_2|M_1, z_1) = \frac{\Delta \delta}{\sqrt{2\pi} (\Delta S)^{3/2}}\left|\frac{dS_2}{d M_2}\right| \exp\left[-\frac{(\Delta\delta)^2}{2\Delta S}\right].
\label{eq:two_barrier_dist}
\eeq
Here $\Delta\delta = \delta_2 - \delta_1$, with $\delta_i = \delta_c/\mathcal{D}(z_i)$, and $\Delta S = S_2 - S_1$, with $S_i = \sigma^2(z_i=0,M_i)$ and the density variance is evaluated using the sharp-$k$ filter.\footnote{Note that since we want to take $z_i$ deep in the MD era, the growth functions 
for adiabatic and isocurvature fluctuations are nearly equal, $\mathcal{D}\equiv \mathcal{D}_{iso}=\mathcal{D}_{adi}$ as discussed in Sec.~\ref{sec:pipeline}.}
The quantity $(df/dM_2) dM_2$ gives the fraction of mass of $M_1$ in microhalos in the mass range $[M_2,M_2+dM_2]$.
Other distributions can be obtained from Eq.~(\ref{eq:two_barrier_dist}) by 
adding appropriate factors (e.g., multiplying by $M_1/M_2$ gives the number distribution of microhalos as a function of their mass at $z_2$). 

In Sec.~\ref{sec:model_examples} we will use Eq.~(\ref{eq:two_barrier_dist}) to show that microhalos make up an $\mathcal{O}(1)$ fraction of the MW DM mass in several representative models, by taking $z_1 = 0$, $M_1 = M_{\mathrm{MW}} \approx 10^{12}M_\odot$, and $z_2 \approx 250$. We again emphasize that this amounts to the crude assumption that all microhalos formed at $z\lesssim 250$ are disrupted.

\section{Probes of Small Scale Structure}
\label{sec:probes}
\subsection{Caustic Microlensing}

An interesting gravitational observable was considered in Refs.~\cite{Diego:2017drh,Oguri:2017ock} and applied to axion miniclusters by Dai and Miralda-Escud\'e~\cite{Dai:2019lud}, and in Refs.~\cite{Arvanitaki:2019rax,Blinov:2019jqc} to other kinds of substructure.
The goal is to observe a background star near a galaxy cluster lens caustic that undergoes microlensing by a compact object in the cluster, 
giving rise to a time-varying light curve with a total magnification that can reach a factor of $10^{3-4}$. During the event duration $\tau$, the star appears to move a distance $d\sim \tau v_{rel} \mu$, where $\mu$ is the magnification and $v_{rel}$ is the star-cluster relative velocity. As the line of sight sweeps over a clumpy dark matter distribution, the magnification varies, leading to jitter in the light curve. These variations may be observable if they are the order of the brightness of the background star when it is not being enhanced by the caustic.

The surface density field $\Sigma$ of a cluster DM halo is related to the line-of-sight integral of the three-dimensional density distribution.
It therefore inherits fluctuations from the intrinsic granularity of the DM distribution in the presence of a high abundance of microhalos. On non-linear scales the density power spectrum 
can be computed in the halo model~\cite{Cooray:2002dia} from the halo mass function $df/d\ln M$ and the Fourier transform of the clump density profile $\tilde\rho^h(q;M)$:
\begin{align}
P_\rho(q) = \bar\rho \int \frac{dM}{M^2}\frac{df}{d\ln M}|\tilde \rho^h(q;M)|^2,
\end{align}
where $\bar\rho$ is a mean density in some region of the cluster.
From this the surface density power spectrum can be estimated by 
partitioning the cluster lens into a series of slices and assuming they have the same mass function, with the result~\cite{Dai:2019lud}: 
\begin{align}
P_\Sigma(q_\perp)=\bar\Sigma  \int \frac{dM}{M^2}\frac{df}{d\ln M}|\tilde \rho^h(q_\perp;M)|^2
\end{align}
where $q_\perp$ is the  surface Fourier mode. Here $\bar\Sigma = \int dL\, \bar\rho(L)$ is the mean surface density.
We are interested in a long line of sight through galaxy clusters, of order a Mpc. In the cases analyzed in~\cite{Dai:2019lud}, the mean surface density along the line of sight is dominated by the cluster, $\bar\Sigma = \bar\Sigma_{cl}$ (as opposed to DM along the line of sight but outside of the cluster).

We will define the sensitivity criterion in terms of the amplitude of fluctuations of the lensing convergence
\begin{align}
\kappa=\Sigma/\Sigma_{crit},
\end{align} 
where
\begin{align}
\Sigma_{crit} &= c^2/(4\pi G D_{eff})\nonumber\\
D_{eff} &= D_L D_{LS}/D_S.
\end{align}
Here $D_{L,S,LS}$ are distances to the lens, the source, and from the source to the lens. $\Sigma_{crit}$ sets the density scale at which an isolated lens  produces multiple images.
The convergence power spectrum is then given by 
\begin{align}
P_\kappa(q)=\frac{P_\Sigma(q)}{\Sigma_{crit}^2},
\label{eq:convergence_ps_def}
\end{align}
or, in dimensionless form,
\begin{align}
\Delta_\kappa^2 = \frac{q^2 P_\kappa}{2\pi}.
\label{eq:dimless_convergence_ps_def}
\end{align}

To obtain rough observational sensitivities we require that $\Delta_\kappa$ is larger than some value $10^{-3-4}$ on some scale $q$ (such that the observed brightness fluctuations are $\mathcal{O}(1)$ if the magnification is $10^{3-4}$). Ref.~\cite{Arvanitaki:2019rax} employed a monochromatic simplified assumption, where one focuses only on clumps of mass $M_s$ composing a fraction $f$ of the dark matter ($df/d\ln M = f M\delta(M-M_s)$ for some constant $0<f<1$). We will see later that the monochromatic assumption provides a good approximation to the convergence power spectrum from a more realistic distribution. In the monochromatic limit and assuming an NFW profile for the clumps, $\Delta_\kappa$ can be expressed analytically,
\begin{align}
\Delta_\kappa = \frac{1}{\log(2/\sqrt{e})}\frac{\sqrt{\Sigma_{cl} f M_s} q r_s g(q r_s)}{\Sigma_{crit} r_s},
\label{eq:monochromatic_convergence}
\end{align}
where $g$ is a function appearing in the Fourier transform of the NFW profile. The quantity $q r_s g(q r_s)$ reaches a maximum of 0.35 at $q r_s=0.77$. In evaluating sensitivities we use  $D_{eff}\sim$ Gpc to compute $\Sigma_{crit}$ and set $\Sigma_{cl}=0.8\Sigma_{crit}$. These values are consistent with the observation of the magnified star LS1 reported in~\cite{Kelly:2017fps}.

To estimate the sensitivity of caustic microlensing, we require that $\max_q \Delta_\kappa(q) > 10^{-3}$ and impose three extra criteria~\cite{Arvanitaki:2019rax}: the lensed light should sweep over many clumps as the star traverses $d$ to enable a statistical treatment; the clumps should be smaller than $d$ to cause significant density fluctuations during the microlensing event; and the clumps should be large enough that the fluctuations they induce in the light curve are not washed out by the finite size of the lensed source star.

The first criterion, that there are many clumps along the line of sight, can be imposed by requiring $f\pi (d/2)^2\Sigma_{cl}/M_s > 10$ or so. This results in a sharp cutoff of $M_s \sim 10^{-2} M_\odot$.  The second criterion, $r_s < d$, can be  imposed by picking representative values for $\tau$, $v_{rel}$, and $\mu$, leading to $d\sim 10^3$ AU~\cite{Arvanitaki:2019rax}. For the third criterion, we take the minimum sensitivity length $\sim 10 $ AU used in Ref.~\cite{Dai:2019lud} and require $r_s > 10$ AU.

\begin{figure}[t!]
\begin{center}
\includegraphics[width=0.5\linewidth]{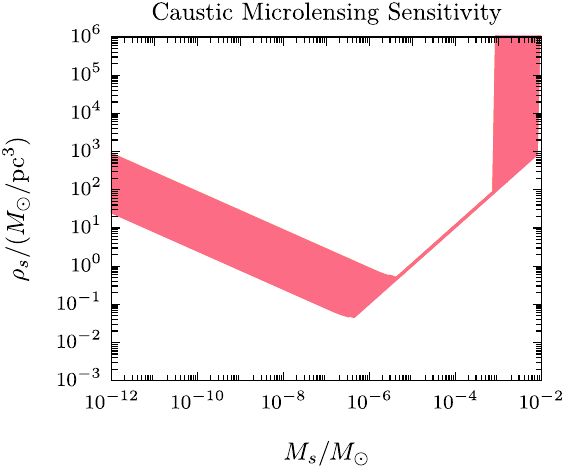}
\caption{Estimate of the sensitivity of caustic microlensing for a monochromatic microhalo distribution in the plane of microhalo scale mass $M_s$ and scale density $\rho_s$ (assuming NFW profiles for these clumps). The bands correspond to 
varying the fraction of DM in microhalos between $0.1$ and $1$. The boundaries of this region are defined in the text. Denser microhalos within these boundaries lead to larger lensing convergence fluctuations.}
\label{fig:causticmicrolensing}
\end{center}
\end{figure}

These criteria and the requirement on $\Delta_\kappa$  lead to the sensitivity curves on the $M_s,\rho_s$ plane shown in Fig.~\ref{fig:causticmicrolensing}.  The thickness of the bands comes from varying $f$ from $0.1$ to $1$. In Sec.~\ref{sec:model_examples} we will show that this monochromatic estimate provides a reasonable approximation to a more realistic calculation of the convergence power spectrum (i.e., using a non-monochromatic distribution) in a few representative models. 

Several phenomena can potentially mimic the lensing signal of enhanced DM substructure. These include the presence of planets in the cluster, blending of the observed star with other faint stars, and surface density fluctuations from gas in the cluster. Ref.~\cite{Dai:2019lud} argued that these processes are unlikely to present a significant background, either because they are rare (e.g., free floating planets and blended stars of comparable brightness to the source), or because they induce fluctuations in convergence on very different length scales compared to microhalos (e.g., cluster gas), or both.

\subsection{Pulsar Timing}

Dark matter clumps can also be probed with searches using pulsar timing arrays (PTAs). This possibility has recently been studied in detail by~\cite{Dror:2019twh,Ramani:2020hdo,Lee:2020wfn}. Two types of signal are possible. The first is a Doppler-shift in the frequency of the pulsar as a clump passes near the pulsar or the Earth. The second type is a Shapiro time-delay if the clump traverses the line-of-sight between the Earth and the pulsar. Depending on how many clumps traverse the vicinity of a pulsar in the array, and how long it takes these signals can be further subdivided into static, dynamic, and stochastic searches. 

While initial work indicated excellent prospects for constraining the properties of clumps, it was realised in Ref.~\cite{Ramani:2020hdo} that some of the DM signal is absorbed by the fit of the pulsar timing model. Incorporating this effect leads to nearly an order of magnitude decrease in the size of the Doppler signal, and a factor of around 4 in the Shapiro signal. The prospects for detection at SKA become  more challenging, and so~\cite{Ramani:2020hdo,Lee:2020wfn} also discuss the capabilities some more futuristic PTAs. The most important parameters in the analysis of~\cite{Dror:2019twh,Ramani:2020hdo,Lee:2020wfn} are the PTA  observation period $T$, timing residuals $t_{RMS}$, cadence $\Delta T$, distance $d$, and number of pulsars $N_P$. The optimistic parameter sets  considered in~\cite{Dror:2019twh,Ramani:2020hdo,Lee:2020wfn} primarily involve decreasing the timing residual to order 10 ns and increasing the number of pulsars in the array to order 1000. There are also increases in the observation time to 30 years and decreasing the observation cadence to 1 week.

Current PTAs include the Parkes Pulsar Timing Array~\cite{Reardon:2015kba} (20 pulsars), the European Pulsar Timing Array~\cite{Desvignes:2016yex} (42 pulsars), and the NANOGrav 11-year data set~\cite{Arzoumanian:2017puf} (45 pulsars). There is some overlap between the arrays, so only  73 of these are independent. Some of the pulsars were combined in the International Pulsar Timing Array data release 1 (IPTA-dr1)~\cite{Verbiest:2016vem}, which used 49 MSPs rather than 73. The more recent IPTA data release 2~\cite{Perera:2019sca} (IPTA-dr2) includes 65 pulsars.

The Square Kilometre Array (SKA) is a next-generation radio-telescope that will commence construction later in the current decade. Ref.~\cite{Rosado:2015epa} estimates that an SKA PTA could include 200 pulsars with 50~ns timing residual and a cadence of 2 weeks. 
These parameters are consistent with other work~\cite{Keane:2014vja,Stappers:2018gwd} - specifically, that SKA2 could discover up to 3000 millisecond pulsars (MSPs), and that 5-10\% of these MSPs are appropriate for use in timing arrays. Results assuming a 20 year observation period at SKA would therefore appear around 2050.

The limitation on the number of pulsars is set by the total number of MSPs in the MW (estimated to be  around 30,000~\cite{Lorimer:2008se,Lorimer:2012hy}) and the fraction suitable for timing arrays. If the latter is of order 5-10\%, achieving a timing array with 1000 pulsars could require discovering an order one fraction of the MSPs in the galaxy.

The current best timing residuals are around 50~ns. Indeed, there is one pulsar in the NANOGrav-11 dataset with $t_{RMS}=30$~ns, although the average in that array is around 10 times higher. The Parkes telescope can achieve residuals of 100~ns on only a few pulsars. One could scale the Parkes parameters to the recently commissioned FAST telescope, which would lead to the conclusion that FAST could achieve residuals of 1-10~ns on some pulsars~\cite{Hobbs:2014tqa}. However,~\cite{Hobbs:2014tqa} indicates this target will be difficult to reach; a ``realistic'' timing array on FAST is considered to be 50 pulsars with residuals of around 100~ns.

The main challenge to achieving low timing residuals is noise from a variety of sources~\cite{Janssen:2014dka}.  In particular, jitter noise on short time scales due to the intrinsic variability of the shape of individual pulsars is already becoming an issue, although there are proposals that may be able to deal with it. On longer time scales there are timing noise and irregularities (``red noise"), and effects from the interstellar medium (ISM). Scattering off the ISM causes a dispersive delay in the signal. For a static source this can be manageable, but millisecond pulsars tend to have relatively high velocities. Consequently over a 20-30 year observation time this dispersion can have an impact. Whether all of these noise sources can be overcome to allow timing residuals of $\mathcal{O}(10)$~ns is as of yet unclear.

Recently~\cite{Lee:2020wfn} has considered the prospects for the observation of dark matter substructure from a number of different dark matter scenarios, including early matter domination and dark photons which we also consider here. For early matter domination and assuming optimistic PTA parameters reheating temperatures less than 1~GeV can be probed. For dark photon dark matter and optimistic parameters PTAs may be sensitive to  masses less than $10^{-7}$~eV (note that for masses $\lesssim 10^{-7}$ eV inflationary production of dark photons cannot saturate the observed DM relic density due to bounds on the scale of inflation~\cite{Akrami:2018odb}).

\subsection{Encounter Rate}
While the previous two subsections focused on gravitational signatures of microhalos, DM substructure can also impact terrestrial direct detection searches if a non-gravitational coupling to Standard Model particles exists.
If most of the dark matter in the galaxy is bound into clumps that formed at high redshift, then the encounter rate of DM constituents with Earth can be drastically different from $\Lambda$CDM. We first estimate the encounter rate assuming (1) the distribution of the largest free clumps is dominated by a single characteristic collapse redshift $z_c$, and (2) the clump properties are well-approximated by the scaling laws of Sec.~\ref{sec:scalings}.

For a monochromatic (i.e., delta-function-like) microhalo mass distribution the encounter rate is
\begin{align}
\tau \equiv \Gamma^{-1}_\mathrm{enc}=\frac{1}{n\sigma v_{rel}}
\end{align}
where the Earth-microhalo relative velocity is $v_{rel}\sim10^{-3}$, the local microhalo number density is $n=\rho_{DM}/M_*(z_c)$, and the cross section is $\sigma = \pi R_*(z_c)^2$. Putting together the pieces, using Eq.~(\ref{eq:rstar_def}) for $R_*$, and dropping ${\cal O}(1)$ numbers, we can express the rate as
\begin{align}
\tau\approx \frac{M_*(z_c)^{\frac{1}{3}}\rho(z_c)^{\frac{2}{3}}}{\rho_{DM} v_{rel}}.
\end{align}
Here $\rho(z_c)$ and $M_*(z_c)$ are given by Eqs.~(\ref{eq:rhozc}) and (\ref{eq:Mstarz}) respectively, so in this approximation the rate is a function of $z_c$, the exponent $n_1$ characterizing the power law enhancement, and the mass $\bar M$ corresponding to the maximum enhanced scale in the matter power spectrum.

A nontrivial distribution of microhalos masses is easily incorporated 
to give a differential encounter rate
\beq
\frac{d\Gamma_\mathrm{enc}}{d\ln M}
= \frac{\rho_{DM}}{M}\frac{df}{d\ln M} \sigma v_{rel},
\label{eq:differential_encounter_rate}
\eeq
where $df/d\ln M$ can be estimated from EPS as in Eq.~(\ref{eq:two_barrier_dist}), and 
the cross-section $\sigma$ implicitly depends on $M$ as above.

\section{Examples}
\label{sec:model_examples}
\begin{figure}[t!]
\begin{center}
\includegraphics[width=0.47\textwidth]{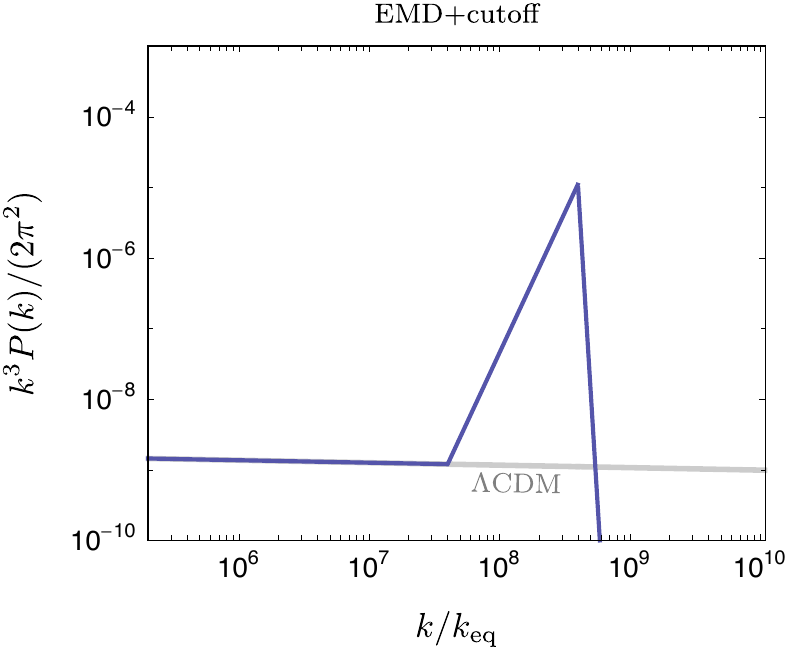}\;\;
\includegraphics[width=0.47\textwidth]{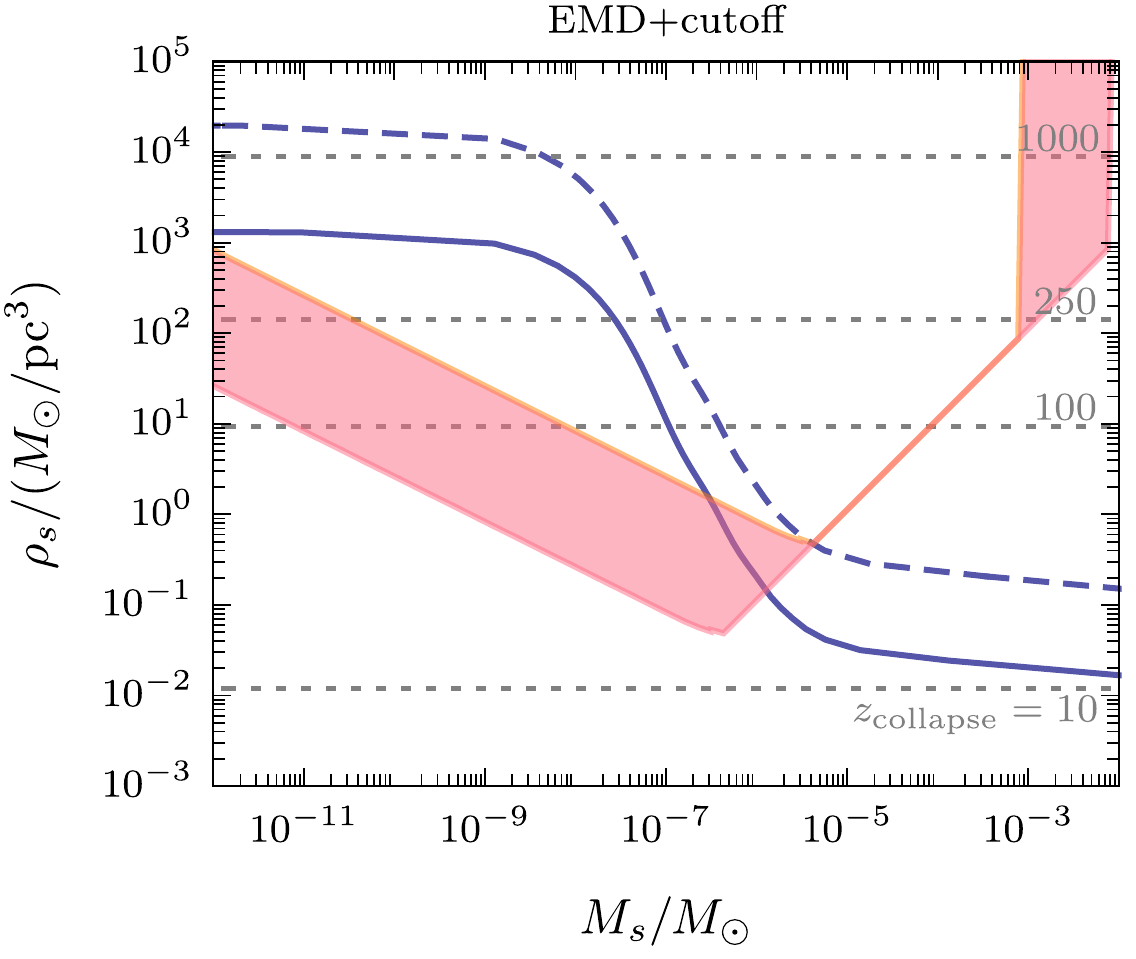}\\
\includegraphics[width=0.47\textwidth]{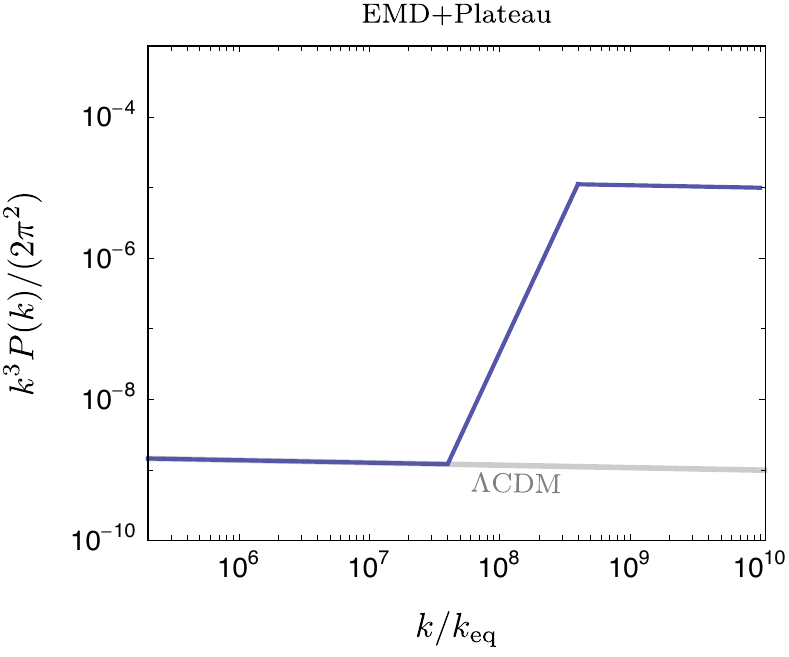}\;\;
\includegraphics[width=0.47\textwidth]{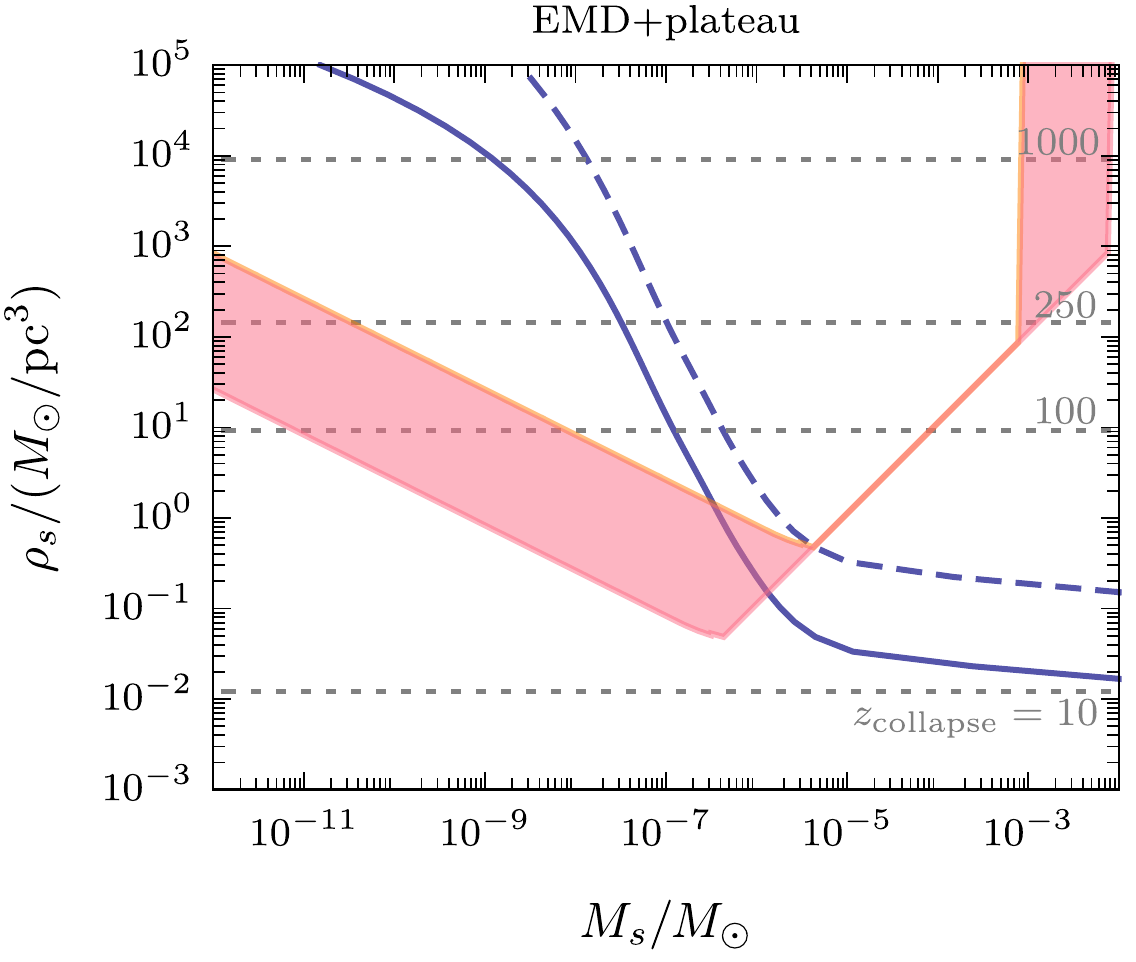}
\end{center}
\caption{Upper left: Example power spectrum resulting from a period of EMD with 10 MeV reheating temperature and a sharp cutoff on small-scale power. Such a cutoff could arise, for example, from the Jeans length in an ALP dark matter model. In this case the corresponding ALP mass is about a nano-eV. Upper right: The $\rho_s-M_s$ relation for this model and the projections for caustic microlensing. The dashed blue curve shows the relation for clumps that form from a 2$\sigma$ fluctuation in the density contrast. Dashed grey lines indicate the collapse epoch $z_c$.
Lower left: Example power spectrum resulting from a period of EMD with 10 MeV reheating temperature and a plateau at small scales. Approximate plateaus arise when any cutoff occurs on much smaller scales than the horizon size at the start of the modified expansion history. Lower right: The $\rho_s-M_s$ relation for this model and the projections for caustic microlensing.}
\label{fig:EMDexamples}
\end{figure}

\begin{figure}[t!]
\begin{center}
\includegraphics[width=0.47\textwidth]{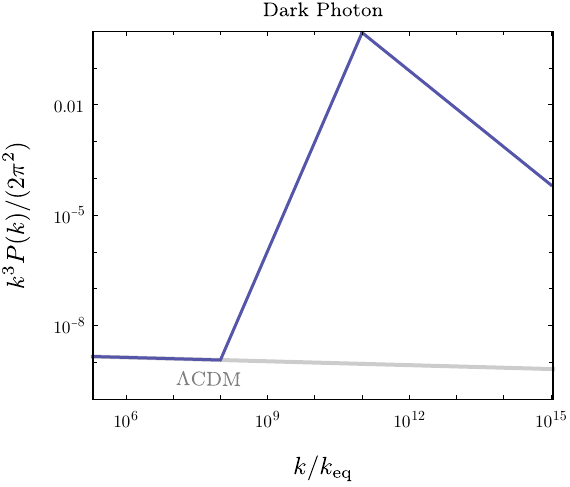}\;\;
\includegraphics[width=0.47\textwidth]{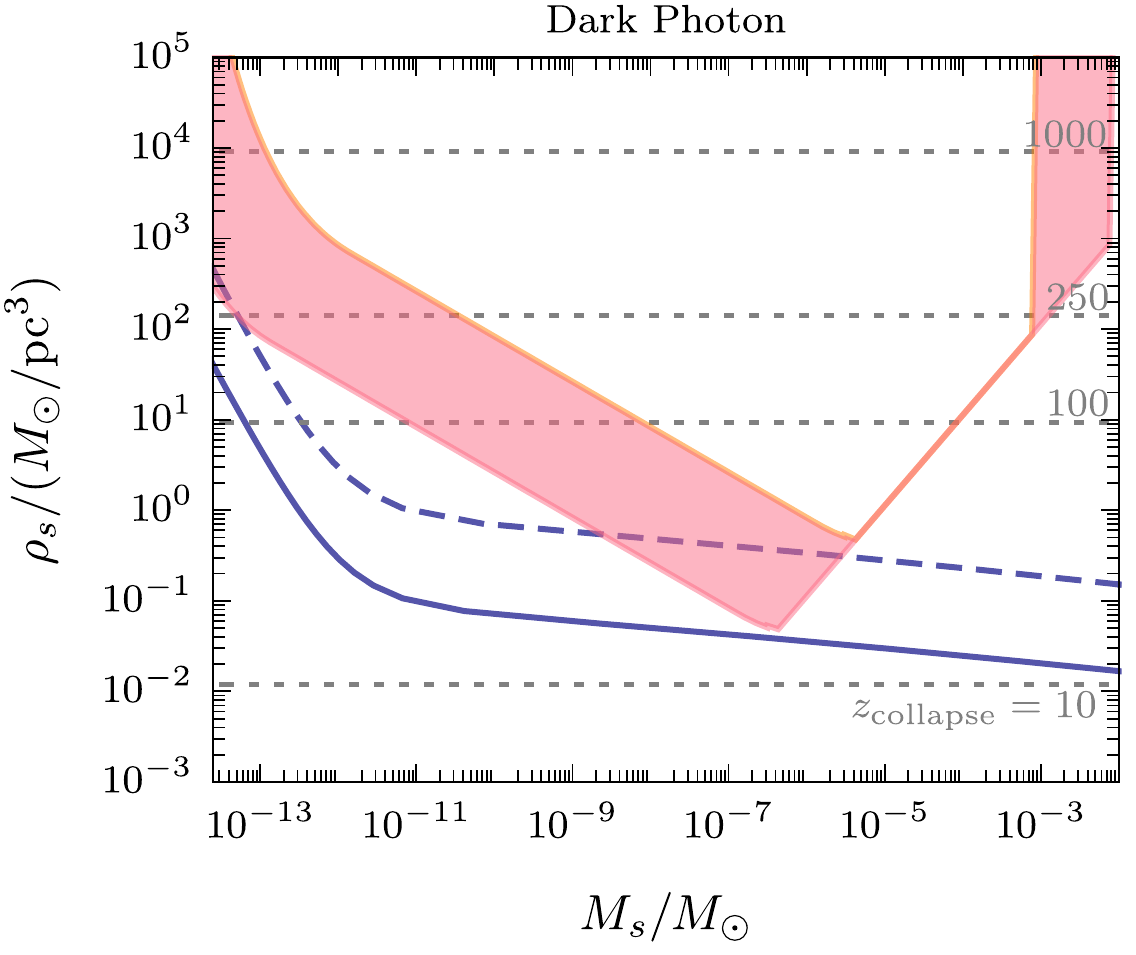}
\end{center}
\caption{Left: Example power spectrum for $10^{-5}$ eV dark photon dark matter produced during inflation. Right: The $\rho_s-M_s$ relation for this model and the projections for caustic microlensing. The dashed blue curve shows the relation for clumps that form from a 2$\sigma$ fluctuation in the density contrast. Dashed grey lines indicate the collapse epoch $z_c$.}
\label{fig:darkphotonexample}
\end{figure}

\begin{figure}
    \centering
    \includegraphics[width=0.47\textwidth]{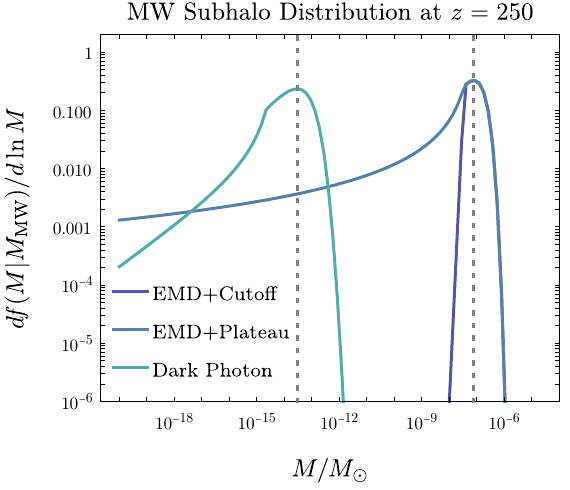}
    \includegraphics[width=0.47\textwidth]{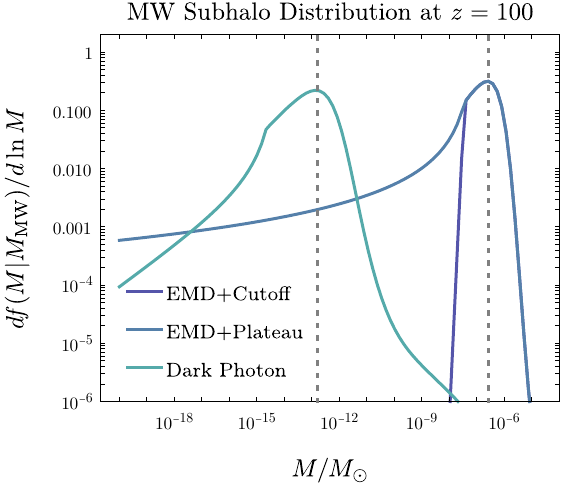}
    \caption{Sample subhalo distribution functions of microhalos in 
   identified at $z_2 = 250$ (left panel) or $z_2 = 100$ (right panel) that end up in a MW-mass galaxy by $z_1 = 0$. The three curves correspond the EMD 
   and dark photon scenarios described in Sec.~\ref{sec:model_examples}. The vertical 
   dashed lines give the characteristic mass $M_*(z_2)$ in each model (defined in Eq.~\ref{eq:mstar_def}). The two EMD examples have $\TRH \approx 10\;\MeV$, while 
   the dark photon curves correspond to $m_{A'} \approx 10^{-5}\;\eV$.}
    \label{fig:EPS}
\end{figure}

\begin{figure}
\includegraphics[width=0.47\textwidth]{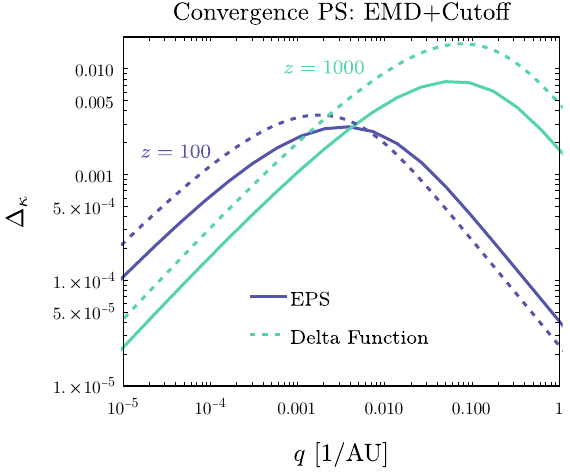}
\includegraphics[width=0.47\textwidth]{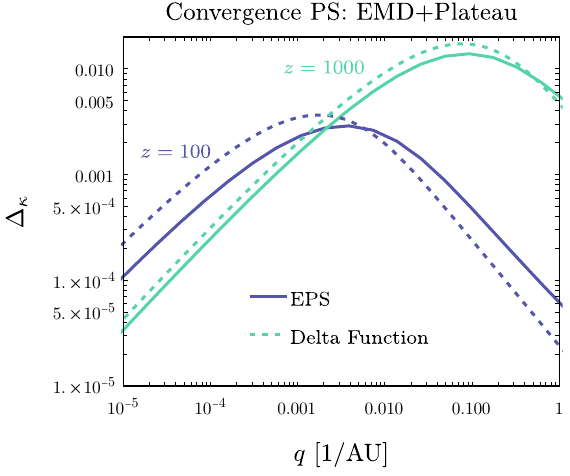}
\caption{Convergence power spectra for the two EMD models (with a high-$k$ cutoff, left, and without, right) with $\TRH \approx 10\;\MeV$, evaluated using the subhalo distributions at $z_2 = 100$ and $z_2=1000$. 
Larger reheating temperatures shift the spectra to larger wavenumbers $q$.
The solid and dotted curves assume an Extended Press-Schechter (EPS) and Delta Function subhalo mass distributions respectively. Convergence fluctuations with a magnitude $\Delta_\kappa \gtrsim 10^{-4}-10^{-3}$ in the range $q\in(10^{-4},0.1)\;\mathrm{AU}^{-1}$ are potentially observable using microlensing of highly magnified stars~\cite{Dai:2019lud}.
\label{fig:convergence_ps_emd}}
\end{figure}

\begin{figure}
\includegraphics[width=0.47\textwidth]{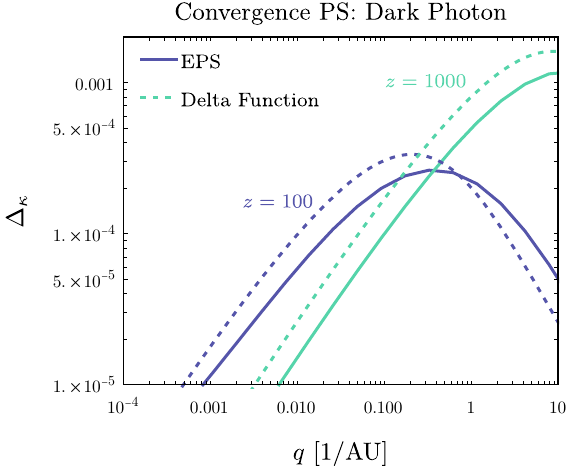}
\caption{
Convergence power spectra for the dark photon model with $m_{A'}\approx 10^{-5}\;\eV$, evaluated using the subhalo distributions at $z_2 = 100$ and $z_2=1000$. The solid and 
dotted curves assume an Extended Press-Schechter (EPS) and Delta Function subhalo mass distributions. 
Larger dark photon masses  shift the spectra to larger wavenumbers $q$.
Convergence fluctuations with a magnitude $\Delta_\kappa \gtrsim 10^{-4}-10^{-3}$ in the range $q\in(10^{-4},0.1)\;\mathrm{AU}^{-1}$ are potentially observable using microlensing of highly magnified stars~\cite{Dai:2019lud}.
\label{fig:convergence_ps_dp}}
\end{figure}

\begin{figure}
    \centering
    \includegraphics[width=0.47\textwidth]{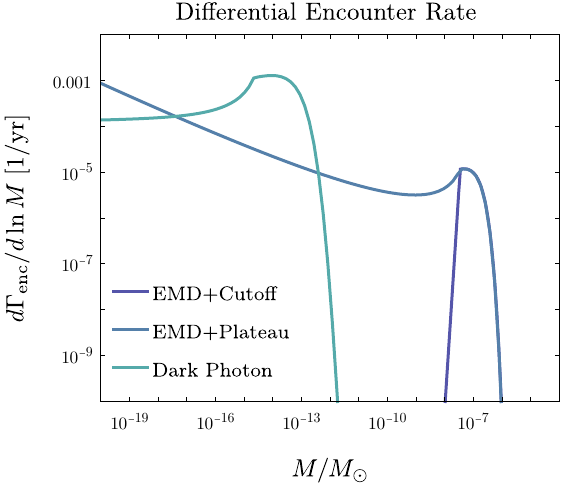}
    \caption{Differential encounter rate as a function of the subhalo mass for EMD-inspired PS enhancements with and without a small-scale cutoff, and the dark photon model described in the text.}
    \label{fig:differential_encounter_rate}
\end{figure}

In this section we apply the results discussed in the 
previous sections to three specific microphysical scenarios: two EMD-like models, and  dark photon DM produced through inflationary fluctuations.
In Figs.~\ref{fig:EMDexamples} and \ref{fig:darkphotonexample}, we show power spectra and quantify the resulting microhalo properties (mass and density), comparing them to the sensitivity of caustic microlensing.
In Fig.~\ref{fig:EPS} we estimate the microhalo 
distribution functions, which are then used to compute lensing convergence power spectra (Figs.~\ref{fig:convergence_ps_emd} and~\ref{fig:convergence_ps_dp}) and Earth-microhalo encounter rates (Fig.~\ref{fig:differential_encounter_rate}).
 Below, we give a brief summary of the models and these results, and then, for illustration, we explain in greater detail the features of early matter domination (arising in the first two models) in subsection~\ref{subsec:emd}.

The first model, denoted ``EMD+cutoff'', exhibits a $k^4$ rise in the power spectrum beginning around $k/\keq\sim 10^7$ and ending in a small-scale cutoff. This power spectrum can be generated by a period of early matter domination with a reheating temperature of around 10 MeV and a Jeans length providing the short distance cutoff. The particular spectrum shown in Fig.~\ref{fig:EMDexamples} corresponds, for example, to an ALP dark matter particle with mass around $10^{-9}$ eV~\cite{Blinov:2019jqc}.\footnote{While the presence of a small scale cutoff is expected due to the Jeans scale, the steepness used here is only a toy model, and it is probably less sharp in a complete treatment. Exponentially falling cutoffs can be achieved, however, in particle DM models with a non-negligible free-streaming scale.} In this case, the modified expansion history allows the relic density to be saturated without fine-tuning the ALP misalignment angle~\cite{Blinov:2019rhb}. The $\rho_s(M_s)$ curve is well within the estimated sensitivity of caustic microlensing for clumps collapsing before $z\sim 50$. 

The second model, denoted ``EMD+plateau," also exhibits a $k^4$ rise in the power spectrum, followed by a plateau. In this case, the transition from $\Lambda$CDM to $k^4$ again corresponds to EMD with a reheating temperature of around 10 MeV. The transition from $k^4$ to the plateau corresponds to the onset of EMD, prior to which the universe is assumed to be radiation dominated. (Logarithmic growth during RD is neglected in the toy model.) A short distance cutoff is typically present at higher $k$, but we extend the plateau to high scales to maximize the difference with the EMD+cutoff model. An example power spectrum of this kind arises in the cannibal models studied in~\cite{cannibalII}. The large scale behavior of the $\rho_s(M_s)$ curve is the same as that of the cutoff model, while the greater power at small scales leads to higher densities at low masses, further strengthening the lensing sensitivity.

The third model, denoted ``Dark Photon" and shown in Fig.~\ref{fig:darkphotonexample}, exhibits a $k^3$ rise in power starting around $k/\keq\sim10^8$ and reaching ${\cal O}(1)$ before descending in a shallow $k^{-1}$ cutoff. This corresponds to dark photon DM with mass $\sim 10^{-5}$ eV produced near the end of inflation~\cite{Graham:2015rva}.\footnote{This is approximately the lowest dark photon mass for which inflationary production can saturate the observed relic abundance of DM given current bounds on the scale of inflation.} Although the boost in power is much larger than in the previous scenarios, it is isocurvature and occurs at smaller scales. Resultingly, the $\rho_s(M_s)$ curve skirts the caustic microlensing projection, only exhibiting significant enhancement below $10^{-11} M_\odot$. This suggests more refined analysis of the microlensing sensitivity is warranted.

Fig.~\ref{fig:EPS} shows the Milky Way subhalo mass distribution derived from extended Press-Schechter for each of the three models. 
These distributions are evaluated at high redshifts $z_2=250$ and $z_2=100$, in the left and right panels, respectively. These are particularly relevant because the rough estimate of clump survival from stellar encounters depends only on the clump density, which in turn depends only on the redshift of collapse. Therefore, by choosing $z_2\sim {\cal O}(100)$ (and $z_1=0$, $M_1=M_{\mathrm{MW}}$) we obtain an estimate for the DM clump distribution in the galaxy today. This amounts to the coarse approximation that all microhalos forming later than $z_2$ are disrupted, while those that form earlier all survive.

Next, we evaluate 
lensing convergence power spectra. These are shown in Fig.~\ref{fig:convergence_ps_emd} for the EMD-inspired models and in Fig.~\ref{fig:convergence_ps_dp} for the dark photon model. In each panel we show the result of evaluating Eq.~(\ref{eq:dimless_convergence_ps_def}) using the EPS distribution with $M_1 = 10^{15}M_\odot$ and $z_1 \sim 1$ (representative of the clusters where caustic microlensing has been observed) and two different values of $z_2$. Since caustic microlensing is a statistical probe, it is potentially sensitive to density inhomogeneities on a wide range of scales, including sub-substructure whose existence is not captured by the EPS mass function and the use of a smooth NFW profile in Eq.~(\ref{eq:convergence_ps_def}). We therefore show the convergence spectrum for $z_2=1000$ and $z_2=100$ to illustrate its range. We note again that the high-$z$ stellar disruption cutoff relevant for the MW does not apply to the clusters relevant for caustic microlensing.
Ref.~\cite{Dai:2019lud} argued that convergence fluctuations with a magnitude $\Delta_\kappa \gtrsim 10^{-4}-10^{-3}$ in the range $q\in(10^{-4},0.1)\;\mathrm{AU}^{-1}$ are potentially observable. We see that the denser, earlier-forming structures result in larger fluctuation amplitude, but on smaller scales (larger $q$). In each panel of Figs.~\ref{fig:convergence_ps_emd} and~\ref{fig:convergence_ps_dp} we compare the power spectra from realistic microhalo distributions to those from monochromatic distributions centered at $M = M_*(z)$,  as evaluated in Eq.~(\ref{eq:monochromatic_convergence}). We see that the latter provides a reasonable, order-of-magnitude approximation to the more realistic distributions. Comparing the two EMD-like examples in Eq.~(\ref{fig:convergence_ps_emd}), we note minute differences in the large-$q$ behavior of the power spectra for the cutoff and plateau cases. The EMD+Plateau example features more power at large $k$ which translates in a larger abundance of smaller halos (see Fig.~\ref{fig:EPS}), which results in slightly more power at large $q$ in the convergence PS. Note that the power spectra shown in Figs.~\ref{fig:convergence_ps_emd} and~\ref{fig:convergence_ps_dp} depend on microphysical parameters which determine the position of the MPS peak and high-$k$ cutoff; for example, increasing $\TRH$ in the EMD models, or the dark photon mass, shifts the MPS peak to larger $k$, which would also translate the convergence power spectra to higher $q$.

We conclude this section with an estimate of Earth-microhalo encounter rates in our benchmark scenarios. The result of evaluating Eq.~(\ref{eq:differential_encounter_rate}) with the EPS microhalo distribution functions in Fig.~\ref{fig:EPS} (with $M_1 = M_{\mathrm{MW}}$, $z_1 = 0$, and $z_2=250$) is shown in Fig.~\ref{fig:differential_encounter_rate}. We see that enhanced substructure leads to Earth-microhalo encounter rates of $\sim $ once in $10^2$ to $10^5$ years. The higher rates correspond to power spectra with more power at large $k$, leading to the formation of lighter microhalos (which have a large number density for fixed total DM energy density). Since we expect that our treatment of disruption is conservative, true rates are likely to be even smaller, since more DM mass is locked up in somewhat heavier microhalos.

\subsection{EMD-induced Bumps in the MPS}
\label{subsec:emd}
The simplified power spectra in Fig.~\ref{fig:EMDexamples} are motivated by scenarios where the cosmological expansion history is modified by a period of EMD, ending shortly before Big Bang Nucleosynthesis (BBN). Here, for illustration, we sketch the enhancement generated by EMD in greater detail. 

Various models can give rise to an EMD era with significant impact on the DM power spectrum. Relic hidden radiation baths, for example, can come to temporarily dominate the energy density of the universe once the lightest dark particle becomes nonrelativistic \cite{Erickcek:2011us,Zhang:2015era}.  Alternatively, EMD can arise  due to the nonthermal production of unstable states, as in misalignment production of moduli~\cite{Banks:1993en,deCarlos:1993wie}. 
Obtaining enhanced small-scale structure from an EMD era requires DM to be kinetically decoupled from the SM radiation bath following reheating \cite{Erickcek:2011us}.  While this condition is often not satisfied for WIMPs \cite{Fan:2014zua}, it can easily be satisfied in models with nonthermal DM (e.g. \cite{Blinov:2019rhb}), or theories where DM arises from a hidden radiation bath that is thermally decoupled from the SM (e.g. \cite{Zhang:2015era, Erickcek:2020wzd}).

During EMD, perturbations grow linearly with scale factor between horizon entry and reheating, resulting in a total enhancement of a density perturbation by a factor $\arh/\ahor$. Since the scale factor at horizon entry scales as $\ahor\sim 1/k^2$, EMD gives rise to a power spectrum that grows like $k^4$. The largest scales that can receive EMD enhancement depend on  the reheating temperature where EMD terminates. Both BBN and the Cosmic Microwave Background (through $N_\mathrm{eff}$) place lower bounds on this temperature, requiring $\TRH\gtrsim 5$ MeV~\cite{deSalas:2015glj,Hasegawa:2019jsa}. For a given collapse redshift, using Eq.~(\ref{eq:Mstarz}), we see that $M_*\sim \bar k^{-3}\sim \TRH^{-3}$. If $\TRH\sim 10$ MeV -- few hundred MeV, the resulting clumps are in the sensitivity region for caustic microlensing. 

The smallest scales that can receive EMD enhancement are also indirectly affected by $\TRH$.
Modes that go nonlinear and collapse during EMD form halos that are dominantly composed of the metastable particle, and they are obliterated by the reheating process \cite{Blanco:2019eij}.  Thus, for adiabatic perturbations that begin at the level of one part in $10^5$, there is an effective upper limit on the amplitude of the EMD-induced enhancement of $10^5$ for modes that enter the horizon during EMD, or $10^4$ for modes that enter during an earlier period of RD (since modes grow by a factor of $\sim 10$ upon horizon entry during RD).

 The location of the peak is primarily determined by the small-scale cutoff, $k_{cut}$.   Here there are a range of possibilities, depending on the microphysics of both DM and the physics generating the EMD era.  
 For example, free-streaming, collisional damping,  and diffusion damping are all typically modeled with a Gaussian cutoff, $T(k)\propto \exp(-k^2/k_{cut}^2)$ \cite{Green:2005fa,Bertschinger:2006nq}.  For free-streaming, $k_{cut}$ is given by the particle free-streaming horizon, $k_{cut}^{-1} = \lambda_{FS}$; for collisional damping, the cutoff is determined by $m_{\phi}$ if DM is predominantly coupled to $\phi$, where $\phi$ is the metastable species responsible for generating EMD \cite{Zhang:2015era, Blanco:2019eij} (i.e., $k_{cut} = a(T=m_\phi) H(T=m_\phi)$ in \cite{Blanco:2019eij}); if DM is instead coupled to the SM radiation bath the temperature of kinetic decoupling, $T_{kd}$, dictates collisional damping scale~\cite{Fan:2014zua} (i.e.,  $k_{cut} = a(T=T_{kd}) H(T=T_{kd})$).

 For another example, in the case of cannibal self-interactions, the cutoff is a more gradual function of $k$, with an envelope falling off as $T(k)\sim (k_{cut}/k)^{2}$ \cite{Erickcek:2020wzd}.
 
 The nature of the peak -- whether it is a relatively sharp feature or a plateau -- depends on whether $k_i$ is smaller or larger than $k_{cut}$, where $k_i$ is the wavenumber of the mode that enters the horizon at the onset of EMD. If $k_i>k_{cut}$, the feature is typically sharp, and $k_i$ plays little role. If, however, $k_i< k_{cut}$, then modes with $k_{cut}>k>k_i$ enter the horizon during a prior period of radiation domination. They experience logarithmic growth, followed by linear growth during  EMD.  
 The resulting feature in the power spectrum is a broad plateau, rather than a sharp bump, eventually cut off near $k_{cut}$ \cite{cannibalII}.

\section{Discussion}
\label{sec:discussion}

Features in the matter power spectrum arise in many different models of dark matter and early universe cosmology, and, as a result, they yield unique probes of both new  particle physics and the universe prior to BBN. 
We have endeavored to give a simplified, phenomenological presentation of the map from enhancements in the power spectrum to the properties of dark matter clumps at late times and their impact on gravitational observables, particularly  caustic microlensing~\cite{Dai:2019lud}.
In general, detailed features of the primordial power spectrum are smoothed out in late-time observables, so that simple broken power law bumps in the MPS are sufficient to model more physical enhancements. The small-$k$ power law, and whether the perturbations are isocurvature or adiabatic, determines many features of the largest, most observable clumps, if they survive until late times. Increasing the slope of the power law increases the collapse redshift for a given scale, which in turn increases the density and decreases the radius of the resulting collapsed objects. Extending the power law to smaller $k$ results in larger, more massive, less dense clumps. If the power law extends to large enough $k$, neither the peak nor the cutoff strongly affects observable prospects.

For more localized enhancements, the details of the peak and cutoff affect the properties of the densest clumps. 
These details are important in some cases and not in others. For example, for caustic microlensing, differences in the MPS cutoff slope lead only to minute differences in the shape of the convergence power spectrum (its amplitude is also modified, but this is degenerate with the fraction of DM in microhalos). In fact, in our  treatment this observable appears to be well approximated using a monochromatic microhalo mass distribution which only depends on the location of the MPS peak and its falloff at small $k$. In contrast, Earth-microhalo encounter rates can be sensitive to the very small-scale behavior of the MPS, strongly influencing direct detection prospects. This is  because shallower slopes in the MPS beyond the peak lead to a large abundance of light microhalos, thereby enhancing the probability of an encounter.

Our analysis relies on a number of assumptions and simplifications which provide numerous directions for future development. We conclude by enumerating some of these points. 
We have only considered clumps that collapse after matter-radiation equality. However, there are other possibilities which lead to structures forming at earlier epochs. One way this can happen is collapse during the radiation-dominated period that precedes matter-radiation equality. The properties of these clumps depend on the free-streaming length of the dark matter, and hence precisely how cold it is. Alternatively, there may be an epoch of non-standard cosmological evolution such as early matter domination (EMD), when structures may also collapse. Clumps that collapse during EMD may disappear at reheating due to the decay of the field responsible for the EMD. This introduces an aspect of model-dependence which is not captured by our phenomenological approach.  A detailed study of structure formation where both possibilities occur was performed in Ref.~\cite{Blanco:2019eij}.

We have focused on clumps whose masses lie within a few orders of magnitude of the mass of the Earth. Partly this is because that is the regime of maximum sensitivity for caustic microlensing. Pulsar timing searches also have sensitivity in this region and at higher masses. It is unclear what searches would be sensitive for clumps with masses substantially less than this.  The existence of lower mass clumps is also more sensitive to UV physics and the mechanism which determines the relic density. For example, ALP clumps from a period of EMD have a lower bound on their masses associated with the size of the mass inside the horizon when the ALP field starts oscillating~\cite{Blinov:2019jqc}.

To predict the properties of the clumps we rely on the Press-Schechter formalism. This is based on assumptions of Gaussian fluctuations, which may not hold for the large enhancements of the power spectra we consider. 
The Press-Schechter picture tells us that larger halos are formed by the hierarchical merges of smaller clump halos. The detailed treatment of this process, and the ultimate fate of the clumps involved requires N-body simulations which are particularly challenging due to the large hierarchies of scales involved. Consequently in our calculations of microlensing sensitivity we assume either a ``monochromatic'' microhalo mass function or one derived from a high-redshift snapshot of an extended Press-Schechter mass function; the true late-time distribution is certainly much more complex. We assume an NFW-type profile for the clumps. Studies of axion miniclusters~\cite{Eggemeier:2019khm} and ultra-compact microhalos~\cite{Delos:2017thv,Delos:2018ueo} both support NFW (or slightly-cuspier-than-NFW) profiles for the first forming halos.

Our projection for the caustic microlensing sensitivity in Fig.~\ref{fig:causticmicrolensing} is quite simplified, and based on~\cite{Arvanitaki:2019rax,Blinov:2019jqc}. We have assumed the existence of a single ``typical" cluster for our projections. This could be improved by considering an ensemble of clusters whose properties are already known. We also have not considered the impact of astrophysical uncertainties and potential backgrounds inherent in the cluster, such as clumps of gas, although these effects have been argued elsewhere to be unimportant~\cite{Dai:2019lud}.

Finally, we do not address the inverse problem. The extent to which any details of the clump mass distribution and profiles (and the early universe physics responsible) can be reconstructed from extended observations using future telescopes is an interesting question which we leave for future work.

\begin{acknowledgments}
  We thank Adrienne Erickcek and Antonella Palmese for useful conversations.
This manuscript has been authored by Fermi Research Alliance, LLC under 
Contract No. DE-AC02-07CH11359 with the U.S. Department of Energy, Office of Science, Office of High Energy Physics. MJD is supported by the Australian Research Council. The work of PD is supported by the US Department of Energy under Grant No. DE-SC0015655.  The work of JS is supported by the US Department of Energy under Grant No. DE-SC0017840.
\end{acknowledgments}

\appendix 
\section{Window Function Dependence}
\label{sec:wf}
The semi-analytical results in the body of the paper depend on the 
choice of the window function $W$ used to filter the density contrast -- see the 
definition of the density variance $\sigma^2$ in Eq.~(\ref{eq:density_variance_def}). All subsequent results, such as typical collapsing masses $M_*$ (Eq.~(\ref{eq:mstar_def})) and the subhalo distribution function depend on this choice. In certain cases, 
the window function choice is physically or theoretically motivated: the real space top hat filter scale is straightforward to interpret in terms of a physical length/mass scale~\cite{Lacey:1993iv}, while a Fourier top hat filter simplifies calculations in the extended Press-Schechter formalism~\cite{Bond:1990iw}. In the context of sharply-rising dimensionless power spectra considered here, the position-space top hat can lead to divergences when computing $\sigma$, while the Fourier top hat always yields finite results. It is therefore important to check 
whether different choices of window functions can change our conclusions. 
In this Appendix we show that this is not the case -- qualitatively similar results are obtained for many different choices of $W$. 

We consider three different window functions: a spherical top hat (TH) with a Gaussian cut off at large $k$, a Fourier space top hat (FTH) and a Gaussian (G):
\begin{subequations}
\begin{align}
W_{\mathrm{TH}}(kR) & = \frac{3(\sin kR - kR \cos kR)}{(kR)^3} e^{-\alpha (k R)^2/2}\\ 
W_{\mathrm{FTH}}(kR) &= \Theta(1 - kR) \\
W_{\mathrm{G}}(kR) & = e^{-(kR)^2/2},
\end{align}
\label{eq:wf_defs}%
\end{subequations}
where we take $\alpha = 10^{-4}$ following Ref.~\cite{Erickcek:2011us}. The mass scale corresponding to $R$ is then given by $M = (4\pi/3,6\pi^2,(2\pi)^{3/2}) \times \bar\rho_{m,0} R^3$ for TH, FTH and Gaussian filters, respectively.\footnote{The equality is only approximate for TH with a Gaussian cutoff. Moreover, the 
normalizations of the FTH and Gaussian filters are ambiguous -- see, e.g.,  Refs.~\cite{Lacey:1993iv,2013MNRAS.428.1774B,Schneider:2013ria} for discussions of these issues.}

\begin{figure}
    \centering
    \includegraphics[width=0.47\textwidth]{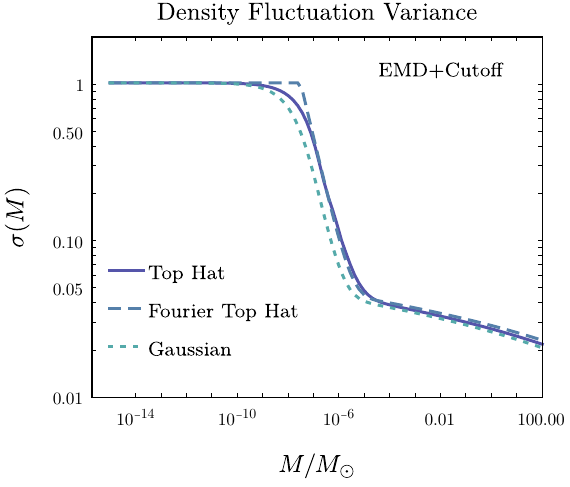}
    \includegraphics[width=0.47\textwidth]{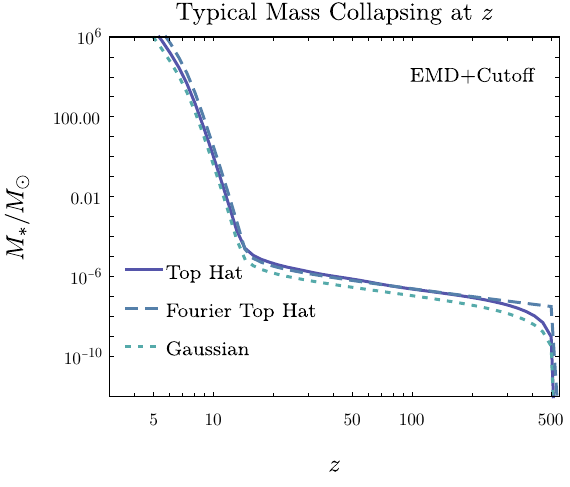}\\
    \includegraphics[width=0.47\textwidth]{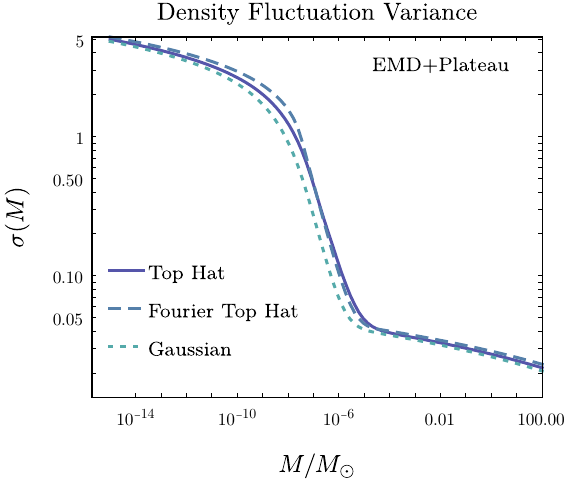}
    \includegraphics[width=0.47\textwidth]{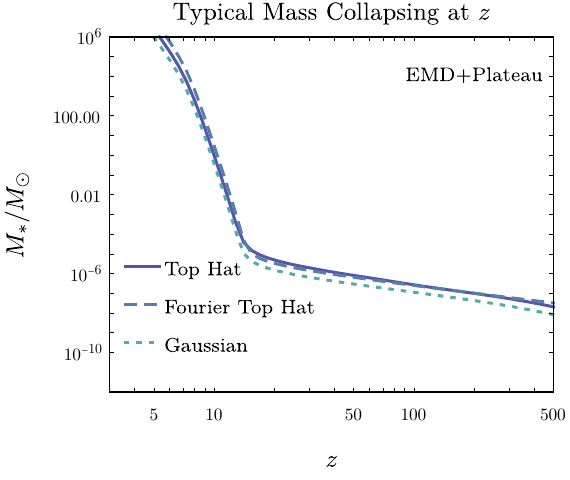}
    \caption{Impact of the window function choice on density fluctuation variance (left column, defined in Eq.~\ref{eq:density_variance_def}) and typical mass collapsing at $z$ (right column, defined in Eq.~\ref{eq:mstar_def}). The two 
    rows correspond to Early Matter Domination-inspired models described in Sec.~\ref{sec:model_examples}, with and without a sharp cutoff at small scales. The solid, dashed and dotted lines correspond to the Top Hat, Fourier Top Hat and Gaussian window functions defined in Eq.~\ref{eq:wf_defs}. In the left column we have fixed $z=1000$.
    \label{fig:wf_variations}
    }
\end{figure}

In Fig.~\ref{fig:wf_variations} we show the result of varying $W$  for two representative MPS enhancements inspired by early matter domination: one has a sharp cutoff at $k=k_\mathrm{peak}$, and another a plateau for $k>k_\mathrm{peak}$. These are described in Sec.~\ref{sec:model_examples}. Different choices of $W$ give estimates of the fluctuation variance and typical collapse mass that differ at most by a factor of $\sim 2-3$. Moreover, the largest differences occur where our model of the MPS enhancement sharply changes slopes; in all physical examples we are aware of the cutoff is much smoother, so we expect that these larger discrepancies to be absent for more realistic power spectra. 
We therefore conclude any choice of window function provides a viable estimate of the density field to our desired accuracy.

\bibliographystyle{JHEP}
\bibliography{biblio}
\end{document}